%% file: SkyrmionReshuffler.tex
\documentclass[11pt,a4paper]{article}

\input{Sections/definitions}

\begin{document}

\title{Skyrmion Gas Manipulation for Probabilistic Computing}
\author[1]{D.~Pinna\thanks{dpinna@protonmail.com}}
\author[1]{F.~Abreu Araujo}
\author[2]{J.-V. Kim}
\author[1]{V.~Cros}
\author[2]{D.~Querlioz}
\author[3]{P.~Bessiere}
\author[3]{J.~Droulez}
\author[1]{J.~Grollier}
\affil[1]{Unité Mixte de Physique CNRS, Thales, Univ. Paris-Sud, Université Paris-Saclay, 91767 Palaiseau, France}
\affil[2]{Centre for Nanoscience and Nanotechnology, CNRS, Univ. Paris-Sud, Université Paris-Saclay, 91405 Orsay, France}
\affil[3]{Institut des Systèmes Intelligents et de Robotique, Université Pierre et Marie Curie, Paris, France}

\renewcommand\Authands{ and }
\date{}
\maketitle

\begin{abstract}
\input{Sections/abstract}
\end{abstract}

\section{Introduction}
\input{Sections/intro}
\section{Motivation: Stochastic Computing}
\input{Sections/stoccompback}

\section{Modelling the Dynamics of Interacting Skyrmions}
\input{Sections/model}
\section{Skyrmion Reshuffler}
\input{Sections/reshuffler}
\section{Skyrmion Neuron}
\input{Sections/neuron}
\section{Conclusion}
\input{Sections/conclusion}
\section*{Acknowledgments}
\input{Sections/acknowledgements}
\appendix
\section{Methods}
\input{Sections/appA-methods}
\section{Qualitative influence of dipole effects on skyrmion repulsion in the ultrathin film limit.}
\input{Sections/appB-1Drep}
\section{Skyrmion properties under different material conditions.}
\input{Sections/appC-Table}

\bibliographystyle{unsrt}
\bibliography{mybib}

\end{document}

%% file: Sections/definitions.tex
\usepackage{graphicx}
\usepackage[caption=false,labelfont=bf]{subfig}
\usepackage{dcolumn}
\usepackage{bm}
\usepackage{comment}
\usepackage{color}
\usepackage{amsmath}
\usepackage{accents}
\usepackage[utf8]{inputenc}
\usepackage[T1]{fontenc}
\usepackage{authblk}
\usepackage{bm}
\usepackage{float}
\usepackage{breqn}
\usepackage[table,xcdraw]{xcolor}

\newcommand{\be}{\begin{equation}}
\newcommand{\ee}{\end{equation}}
\newcommand{\bea}{\begin{eqnarray}}
\newcommand{\eea}{\end{eqnarray}}
\newcommand{\atan}{\mathrm{atan}}

%% file: Sections/abstract.tex
The topologically protected magnetic spin configurations known as skyrmions offer promising applications due to their stability, mobility and localization. In this work, we emphasize how to leverage the thermally driven dynamics of an ensemble of such particles to perform computing tasks. We propose a device employing a skyrmion gas to reshuffle a random signal into an uncorrelated copy of itself. This is demonstrated by modelling the ensemble dynamics in a collective coordinate approach where skyrmion-skyrmion and skyrmion-boundary interactions are accounted for phenomenologically. Our numerical results are used to develop a proof-of-concept for an energy efficient ($\sim\mu\mathrm{W}$) device with a low area imprint ($\sim\mu\mathrm{m}^2$). Whereas its immediate application to stochastic computing circuit designs will be made apparent, we argue that its basic functionality, reminiscent of an integrate-and-fire neuron, qualifies it as a novel bio-inspired building block.

%% file: Sections/intro.tex
Magnetic skyrmions promise unique opportunities for the processing, storage and transfer of information at the intersection of both spintronics and nanoelectronics~\cite{fert2013skyrmions, nagaosa2013topological, sampaio2013nucleation, iwasaki2013universal, zhang2015magnetic, koshibae2015memory, fert2017magnetic}. The experimental verification of skyrmion spin configurations has been a forefront research topic in magnetism over the past decade. Initially observed in bulk non-centrosymmetric crystal lattices~\cite{neubauer2009topological,pappas2009chiral,tonomura2012real,yu2010real,huang2012extended}, they have been more recently stabilized in ultrathin ferromagnetic nanostructures strongly affected by the Dzyaloshinski-Moriya interaction (DMI) resulting from coupling to a heavy metal substrate~\cite{heinze2011spontaneous,romming2013writing}. In such materials, skyrmions are exceptionally stable spin textures capable of enduring room-temperature environments~\cite{jiang2015blowing,moreau2016additive,woo2016observation,boulle2016room} and manipulable at small current densities ($10^6 - 10^{11}\,\mathrm{A}/\mathrm{m}^2$)~\cite{jonietz2010spin,yu2012skyrmion,woo2016observation,jiang2016direct,hrabec2016current,litzius2017skyrmion}. Furthermore, their topological resilience \cite{bogdanov2001chiral,nagaosa2013topological} allows them to avoid pinning onto defects~\cite{iwasaki2013current,lin2013particle} and guarantees particle number conservation under a wide range of operating conditions~\cite{reichhardt2015collective,muller2015capturing}. 

The appeal of skyrmions is so wide that it has defined a field of its own, skyrmionics, which refers to the emerging technologies based on magnetic skyrmions as information carriers. In an effort to push for skyrmion-based electronics, challenges ranging from their creation and annihilation~\cite{nagaosa2013topological,zhang2015skyrmion, tchoe2012skyrmion,legrand2017room}, the conversion of their topological properties~\cite{zhang2015magnetic, zhou2014reversible}, as well as their efficient transmission and read-out~\cite{nagaosa2013topological,iwasaki2013universal,sampaio2013nucleation,tomasello2014strategy, koshibae2015memory} are being tackled and solved. Recently, proposals have exploited their nano-metric size and high mobility~\cite{muhlbauer2009skyrmion,lin2013particle,yu2012skyrmion, fert2013skyrmions} in the context of skyrmion-based racetrack memories~\cite{sampaio2013nucleation,zhang2015skyrmion,kang2016voltage}, logic gates~\cite{zhang2015magnetic}, voltage-gated transistors~\cite{zhang2015skyrmiontransistor} and synaptic devices~\cite{huang2017magnetic}. All such applications typically constrain the skyrmion motion onto 1-dimensional tracks reminiscent of previous device designs seeking to employ domain wall motion. Very little has been proposed to take advantage of the analogy between skyrmions and a generic free moving particle. Particularly, the full 2-dimensional freedom of skyrmion motion has not been leveraged either at a functional level or as a technique to reduce the surface imprint of proposed devices.

\begin{figure}
	\centerline{\includegraphics[width=4in]{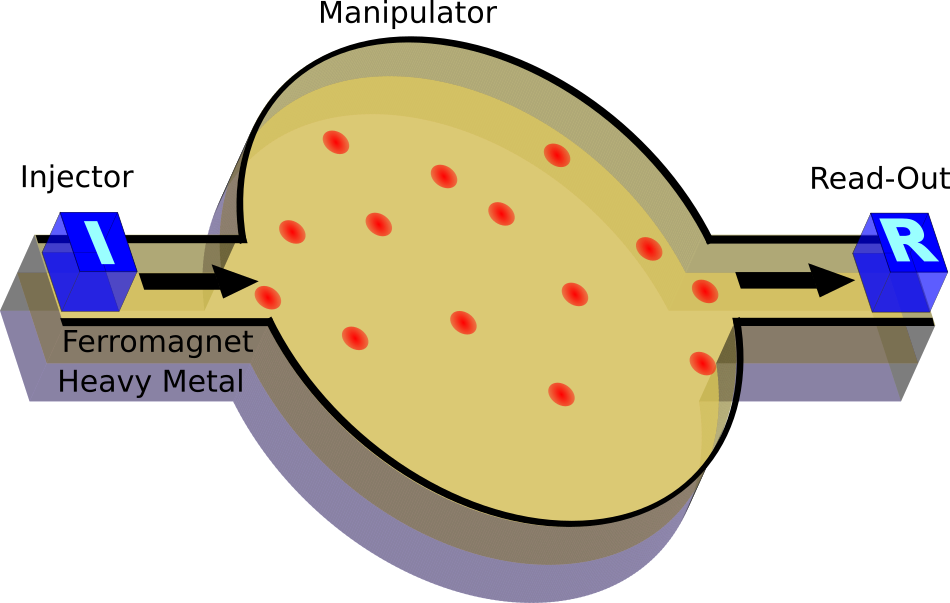}}
        \caption{{\footnotesize Bare-bones schematic of a generic skyrmionic device constructed with a thin film ferromagnet/heavy metal bilayer. One or more skymions are injected into the geometry where manipulation occurs before readout.}}
	\label{fig:schematic}
\end{figure}

All such applications effectively employ geometries where one or more skyrmions are injected, manipulated and read out (see Fig.~\ref{fig:schematic}). Currently, the standard approach for studying their behavior is to employ micromagnetic simulations capable of modeling the nonlinear evolution of the structure's magnetic texture over time~\cite{miltat2007numerical}. Whereas micromagnetic techniques can be very precise, the execution time of the finite-difference method they employ scales rapidly with the volume of the structure itself~\cite{smith1985numerical}. As such, resolving the behavior of large structures and circuits over long time periods becomes impractical even on the nanosecond scale. It is therefore important to have an efficient method for simulating the behavior of skyrmion ensembles which scales with the number of skyrmions instead. Such a technique should be capable of capturing both the influence of magnetic dipole effects as well as boundary interactions on the ensemble dynamics.   

At a device level, skyrmionic applications are not {\it currently} expected to compete significantly with present-day CMOS implementations of boolean logic circuits whose strengths lie in the accuracy and precision of bitwise operations. However, limits in the exponential miniaturization of transistor-based architectures exposes bottlenecks both with regards to speed as well as their power consumption. This is particularly relevant when requiring for hardware to enable software capable of performing fast inference over large amounts of data. As an example, much attention has been directed towards the development of artificial neural networks for the execution of deep-learning techniques~\cite{basheer2000artificial}. Such solutions typically seek to draw inspiration from the probabilistic and massively parallel nature of biological information processing~\cite{colliaux2016cell} whose virtues have been explored for the past half century~\cite{von2012computer}. In this context, we propose that greater focus should instead be given to the use of skyrmions as basic elements in the design of probabilistic device architectures. By leveraging skyrmion number conservation, thermal susceptibility and ultra-low power transport properties, we believe that a fertile route towards disruptive applications can be achieved. Along this line, our work presents two novel devices implementing basic building blocks for probabilistic computing.

Starting from the collective coordinate dynamical theory introduced by Thiele~\cite{thiele1973steady}, this paper develops an N-body framework capable of modelling the dynamics of skyrmion ensembles within their geometries and successively demonstrate how they may be employed to perform probabilistic computing.  To this aim we first, in Section 2, motivate the deep technological challenges inspiring this work by giving a general background on stochastic computing and justifying the need for a compact and energy efficient stochastic signal reshuffler which we propose employing skyrmions as information carriers. In Section 3, we use micromagnetic simulations to characterize the repulsive skyrmion-skyrmion and skyrmion-boundary interactions. These then enter phenomenologically into the N-body Thiele model where the ensemble dynamics become numerically solvable out to long timescales. These advantages permit us in Section 4 to model the Skyrmion Reshuffler leveraging the full freedom of skyrmionic motion. We characterize the device functioning performance and expected energy consumption. Finally, in Section 5, we argue how our skyrmion reshuffler, subject to minor modifications, can serve a much more general purpose by effectively working as an analog integrate-and-fire neuron~\cite{burkitt2006review}. These results lay the groundwork for the application of skyrmionic devices as bio-inspired building blocks for non-conventional computing strategies. 

%% file: Sections/stoccompback.tex
\subsection{Motivation}

\begin{figure}
	\centerline{\includegraphics[angle=-90,width=3in]{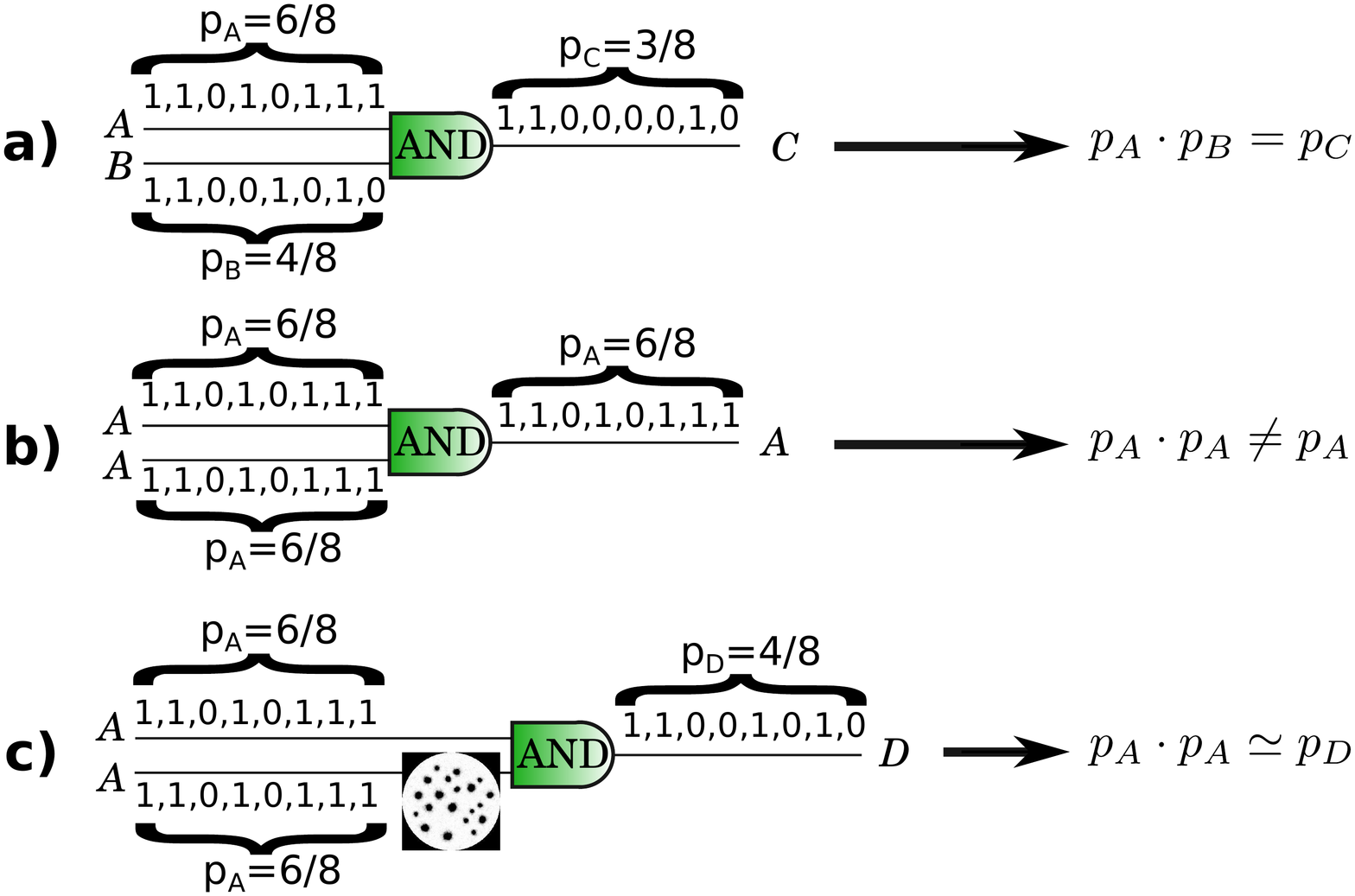}}
	\caption{{\footnotesize a) AND-gate implementation of the multiplication operation under the stochastic computing paradigm. The p-value of the output signal is equal to the product of the input signals' p-values. b) Whenever correlations exist between input signals, the AND-gate doesn't perform the expected multiplication. c) Forcing the input signals through a skyrmionic reshuffler allows correct multiplication operation even in the presence of strong correlations between the input signals.}}
	\label{ANDgates}
\end{figure}

As feature sizes in semiconductor systems are scaled down to the point where their deterministic behavior cannot be guaranteed, the necessity for error-correcting codes~\cite{rao1974error} and hardware redundancies~\cite{pratt2008fine,koren2010fault} have become indispensable for reliable data processing. In this context, {\it stochastic computing} seeks to embrace probabilistic computing elements to bypass these issues by achieving tremendous gains in signal processing efficiency at the cost of degradation in computational precision~\cite{gaines1967stochastic,alaghi2013survey,morro2015ultra,canals2016new,friedman2016bayesian}. Instead of defining operations as precise bitwise manipulations of stored {\it finite} binary numbers, one instead aims to encode numerical values as the probability of seeing a $1$ or $0$ in a random sequence of binary bits ({\it bitstream}, as in Fig.~\ref{ANDgates}~\cite{von1956probabilistic}).  This probability, known as the {\it p-value}, can be also thought as the statistical ratio of up-time to signal length in a random telegraph noise signal. CMOS circuits consisting of cascaded logic gates can then be shown to perform mathematical operations on the p-values of distint random streams \footnote{We will refer to telegraph noise signals and bitstreams interchangeably throughout this manuscript} in real-time. Complex circuits have already been proposed to perform operations such as square-rooting~\cite{toral2000stochastic}, polynomial arithmetic~\cite{qian2008synthesis,li2009reconfigurable,qian2011architecture}, matrix operations~\cite{mars1976high} as well as the "tansig" transform function employed in neural-networks~\cite{zhang2008stochastic}. The output of such operations will result in an output bitstream whose p-value can be read out and possibly operated on successively. Last but not least, stochastic computing circuitry is inherently tolerant to faults such as {\it soft bit-flips}~\cite{baumann2005radiation,parhi2015effect}. Consider the act of randomly flipping one binary bit from a sequence of {\it N} bits (ex: $010001101\to 010011101$): the binary number encoded changes drastically while the p-value encoded is altered only by a factor of $o(1/N)$.

A very attractive property of stochastic computing is that it enables low-energy cost implementations of arithmetic operations using standard logic elements. As an example, consider the effect of feeding two indefinitely long random streams with p-values $p_A$ and $p_B$ into an AND-gate (see Fig.~\ref{ANDgates}a). The AND-gate's properties will produce an output signal with p-value $p_C=p_A\cdot p_B$, thus implementing the stochastic computing equivalent of multiplication. Whereas the result of the multiplication is produced in real-time as the input signals are fed to the AND-gate, the accuracy of the result depends on how much of the output signal one wishes to sample. This trade-off of {\it speed vs accuracy} is central in the development of integrated circuits performing stochastic computing. A detailed description of the initial encoding and tuning/setting of input p-values goes beyond the scope of this work. We note however that, aside from the traditional shift-register based techniques~\cite{golomb1982shift,gupta1988binary}, a compact and energy efficient approach for generating suitable stochastic signals can be achieved through current-biased superparamagnetic magnetic tunnel junctions~\cite{ohta2007large,locatelli2014noise,perricone2016design,vodenicarevic2017low} whose individual use in neuromorphic computing architectures has also been recently explored~\cite{sengupta2016probabilistic,faria2017low,sutton2017intrinsic}. This ties in well with the spintronics-based toolbox that can already be used for the nucleation and read-out of skyrmions~\cite{romming2013writing,jiang2015blowing}, thus allowing for the development of technologically consistent devices~\cite{venkatesan2015spintastic}.

A major issue has however impeded the blossoming of stochastic computing as an industrially viable competitor. Namely, cascading gates operating on non-fully random signals propagate unwanted correlations rapidly even after a few elementary operations~\cite{gupta1988binary,jeavons1994generating,alaghi2013exploiting,moons2014energy,manohar2015comparing,friedman2016bayesian}. In Figure~\ref{ANDgates}b, one can immediately see that feeding two identical copies of the same input signal produces an output signal whose p-value does not represent a product operation in any way. It is therefore crucial to be able to regularly reshuffle signals so that they stay uncorrelated. We call this element a stochastic reshuffler, capable of {\it copying} an input stream into an uncorrelated new one while preserving the original p-value. By prefixing such devices to each gate in a circuit (see Fig.~\ref{ANDgates}c), inconvenient correlations can be washed out effectively.

As of present, no efficient hardware stochastic reshufflers are known to exist. In CMOS, reshuffling signals can be done by combining a pseudo random number generator with a shift register or counter (both requiring long term memories)~\cite{tehrani2010relaxation}. This means that each reshuffling operation has a large area imprint (a typical integrated shift-register has a linear size of $\sim10-100\,\mu\mathrm{m}$), and consumes a lot of energy. It is therefore impossible to insert these reshufflers after each calculation stage in a stochastic computing circuit, preventing the realization of any large scale demonstration of stochastic computing on chip. To overcome these technological obstacles, we propose a novel reshuffler design as the first low-energy, compact device proposal of its kind: the Skyrmion Reshuffler~\cite{pinna:hal-01399341}. 

\subsection{Skyrmion Reshuffler}

Our original concept converts an input bitstream into a sequence of skyrmions whose order can be thermally reshuffled due to the diffusive 2-dimensional dynamical nature of interacting skyrmions at finite temperature. If the order of this sequence can be altered enough by the thermal noise affecting skyrmion motion inside the device geometry, the new skyrmion sequence can then be {\it read} as a new uncorrelated output signal with p-value identical to that of the input signal.

The device (depicted in Figure 3) consists of two circular chambers with input/output conduit tracks capable of ushering skyrmions into and out of the chambers. The net drift of skyrmions is achieved by a static current flowing across the entire structure from one conduit to the other (see Methods). Skyrmions injected can be nucleated either singularly via injection of localized spin-currents with magnetic tunnel junctions (MTJ)~\cite{sampaio2013nucleation}, using in-plane spin-torques on a notch~\cite{iwasaki2013current}, by rapid proliferation with appropriately constructed input conduit geometries~\cite{jiang2015blowing,yu2016room} or via periodic shedding off of material defects with homogeneous DC currents~\cite{everschor2016skyrmion}. Analogously, the read-out elements could consist of MTJs capable of measuring magnetoresistance changes due to the passing of skyrmions through the underlying magnetic layer~\cite{hanneken2015electrical}. We will summarily assume injection/detection capabilities by providing each track with an injection/detection element. 

The up-and-down (binary) states of the bitstream are used to select into which chamber to inject the generated skyrmions (see Fig. 3). We do so by selectively nucleating skyrmions at a constant rate ($r_{\mathrm{inj}}=0.2\;\mathrm{nucleations}/n\mathrm{s}$ in the simulations presented here) onto the input conduit of one of the two chambers ($\mathrm{UP}$ and $\mathrm{DOWN}$) depending on the input signal's state. Doing so generates a total population of $N$ skyrmions proportional to the length $\tau_{\mathrm{input}}$ of the input signal ($N=r_{\mathrm{inj}}\tau_{\mathrm{length}}$) and chamber populations $N_{\mathrm{UP}}=pN$ and $N_{\mathrm{DOWN}}=(1-p)N$ given by the $p$-value of the input signal. The current-induced drift then pushes the skyrmions across their respective chambers toward the output track where they will be detected. The sequence of outgoing skyrmions can then be decodified into an outgoing signal. Whenever an outgoing skyrmion is read-out from the $\mathrm{UP}$($\mathrm{DOWN}$)-chamber we switch the outgoing signal into an {\it up} ({\it down}) state. As a result of this procedure, the output signal's length will be given by the time $\tau_{\mathrm{output}}$ taken to read all the skyrmions and, by construction, its p-value will be identical to that of the input signal as long as the relative skyrmion particle numbers in the chambers are conserved. The basic operating principles of the Skyrmion Reshuffler allow it to also work as a device capable of generating stochastic bitstreams with well-defined p-values. To do so, it suffices to inject skyrmions into the two chambers with a specified ratio and allowing the reshuffling to scramble the input order into a random outgoing skyrmion sequence. 

In order to quantify the performance of skyrmion reshuffling as a function of chamber materials and geometries, we first model the dynamics of thermally diffusing assemblies of interacting skyrmions.

\begin{figure}
	\centerline{\includegraphics[angle=-90,width=4in]{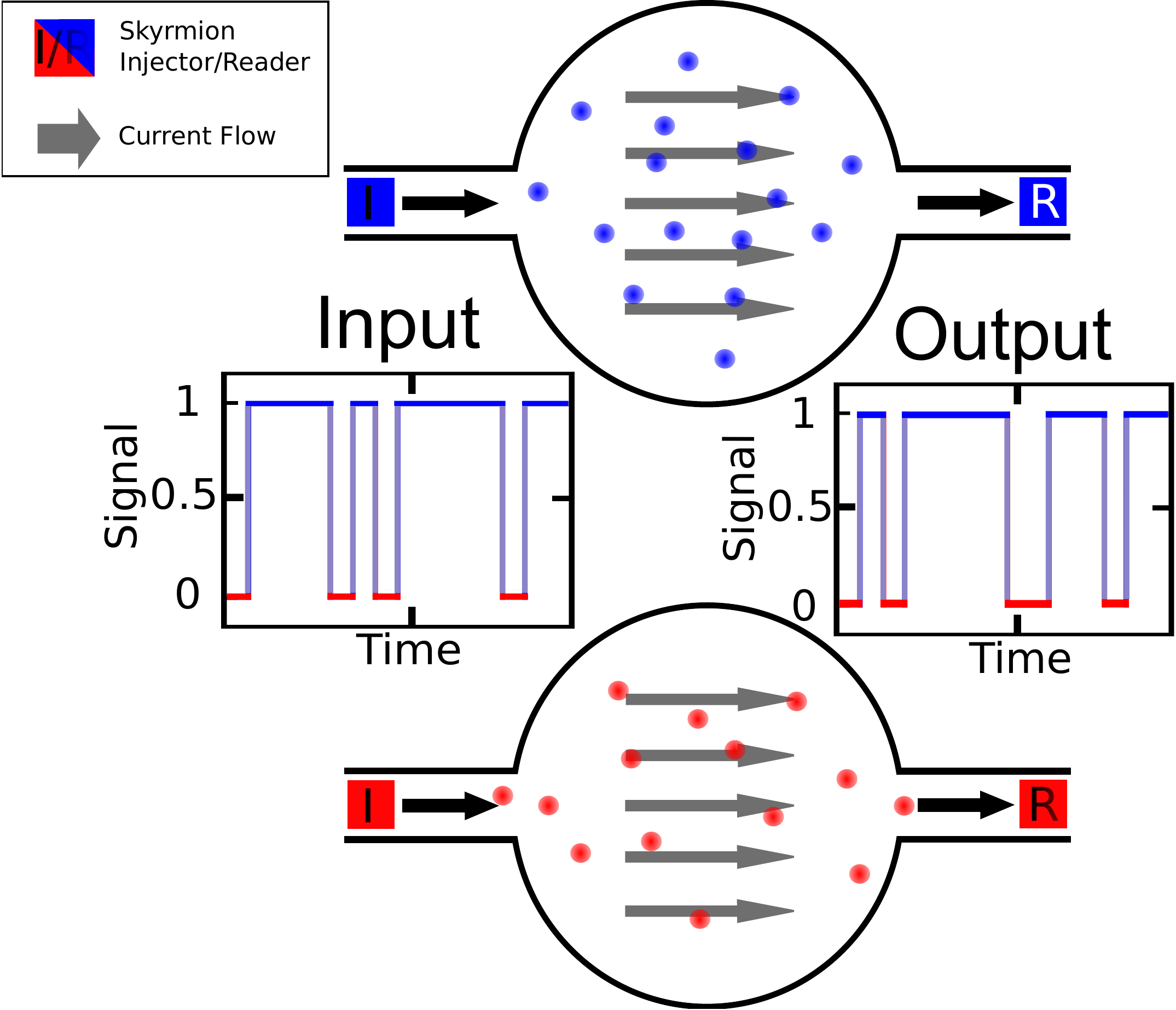}}
	\caption{{\footnotesize The proposed device consists of two magnetic chambers into which skyrmions are injected depending on the state of an input telegraph noise signal. The net drift of the skyrmion particles due to a constant current flow along with the thermal diffusion in the chambers leads to an exit order that can be significantly different from that of entry. This is employed to reconstruct a new outgoing signal with the same statistical properties as the first as well as being uncorrelated from it.}}
	\label{Reshuffler}
\end{figure}

%% file: Sections/model.tex
\subsection{Isolated Skyrmion Dynamics}

The motion of an isolated skyrmion in a two dimensional film can be described by a modified version of Newton’s equation which tracks the coordinate evolution of the skyrmion's center. Assuming translational invariance of the skyrmionic spin texture, the {\it Thiele}-equation of motion can be used~\cite{everschor2011current,clarke2008dynamics,thiele1973steady,iwasaki2013universal}:

\begin{eqnarray}
&\mathbf{\hat{G}}^{\alpha}\cdot\mathbf{\dot{x}} = \mu\mathbf{\hat{G}}^{\eta}\cdot\mathbf{v}_s+\mathbf{F}_{\mathrm{Th}}+\mathbf{F}_{\mathrm{Ext}}\label{eqn:Evo}\\
&\hat{G}^{\sigma}_{ij} = G\,\epsilon_{ij}+\sigma D\,\delta_{ij}\label{eqn:G}
\end{eqnarray}

where $\mathbf{\hat{G}}$ is the gyrotropic matrix, $\epsilon_{ij}$ the Levi-Civita tensor, $\delta_{ij}$ a Kr\"{o}necker delta, and $\sigma$ in (\ref{eqn:G}) is equal to either the damping constant $\alpha$ (when referring to the left side of (\ref{eqn:Evo})) or a parameter $\eta$ characterizing both the net ratio of adiabatic and non-adiabatic spin-torque intensities as well as the ratio of damping-like to field-like torques depending on the geometries of current injection. Doing so, our model is general enough to capture equally well the skyrmion driving motion originating from both spin-torques due to in-plane currents as well as spin-orbit torques arising from the underlying heavy metal layer~\cite{tretiakov2007vortices,lin2013particle,ado2017microscopic}. Lastly, the parameter $\mu$ models the net intensity due to the combination of these two driving effects. Time is in units of $\gamma M$, where $\gamma$ is the gyromagnetic ratio and $M$ the magnetic texture's local moment magnitude. Under the assumption of an invariant skyrmion profile, the gyrotropic matrix (\ref{eqn:G}) is composed by the gyrovector $G$, a topological  invariant arising from the {\it twist} in the spin-texture, and the dissipative diadic $D$ (also known as the {\it gyrodamping}) which, together with $\alpha$, characterizes the net friction acting on the skyrmion~\cite{tretiakov2007vortices,clarke2008dynamics,tveten2013staggered,schutte2014inertia,thiele1973steady}. Both are typically computed explicitly from the static magnetic profile:

\begin{eqnarray}
G &=& \int\mathrm{d}\mathbf{r}\,\mathbf{n}\cdot\left(\partial_x\mathbf{n}\times\partial_y\mathbf{n}\right)\label{eq:G}\\
D &=& \int\mathrm{d}\mathbf{r}\,\left(\partial_x\mathbf{n}\cdot\partial_x\mathbf{n}+\partial_y\mathbf{n}\cdot\partial_y\mathbf{n}\right)/2,\label{eq:D}
\end{eqnarray}
where $\mathbf{n}$ is the local, unit-normalized, magnetization orientation. A very convenient property of both terms is that they do not scale with the skyrmion's size as long the skyrmion is larger than a few times the domain wall width $\sqrt{A_{ex}/K_u}$ (where $A_{ex}$ and $K_u$ are the material exchange constant and perpendicular magnetic anisotropy respectively). This ensures that their values computed in zero-temperature micromagnetic simulations will also be valid when modelling finite-temperature skyrmions. Thermal effects are known in fact to excite breathing modes in the skyrmion profile which imply a periodic fluctuation in the skyrmion size~\cite{kim2014breathing}. These are expected to qualitatively alter the dynamics of large skyrmions ($\sim 200 n\mathrm{m}$) where second order inertial terms should be included in the respective Thiele model~\cite{schutte2014inertia,buttner2015dynamics}. For the physical parameters chosen in our simulations (see Methods), we find stable isolated skyrmions with a typical diameter of $\sim60 n\mathrm{m}$. 

The three force terms appearing on the right side of (\ref{eqn:Evo}) model various interactions that dynamical skyrmions are subject to. Both the conduction of electrons through the spin-texture and the pumping of spin through spin-orbit-torque effects are known to induce both an adiabatic and non-adiabatic spin-torque~\cite{tatara2008microscopic} (usually refered as damping-like / field-like spin-torques when spin orbit torques are considered) giving rise to well-defined force components acting on the skyrmion profile~\cite{schulz2012emergent, everschor2011current}. These current-induced drift effects are captured by the first term where $\mathbf{v}_s$ is the spin-drift velocity (directly proportional to current density $j$). In the special case where $\eta=\alpha$, skyrmions drift along the current flow lines. In all other circumstances, however, the discrepancy between $\alpha$ and $\eta$ leads to current-induced skyrmion flows which proceed at an angle to the current-flow driving them~\cite{sampaio2013nucleation,jiang2016direct}. 

Similarly, the thermal forces are modelled by the collective action of independent random magnetic fields acting on the texture's local magnetic moments. The net effect results in an additive, homogeneous zero-mean stochastic term to the Thiele dynamics~\cite{mertens2000nonlinear,troncoso2014brownian}:

\begin{eqnarray}
\label{eqn:thielenoise}
\langle\mathbf{F}_{\mathrm{Th}}\rangle &=& 0\\
\langle F_{\mathrm{Th},i}(t) F_{\mathrm{Th},j}(t')\rangle &=& 2\alpha D\frac{k_BT}{\gamma M}\delta_{i,j}\delta(t-t'),\label{eqn:diffconst}
\end{eqnarray}
where $k_BT$ is the thermal energy and $\langle\cdot\rangle$ represents averaging over noise realizations. From~(\ref{eqn:Evo}), the mean squared displacement of skyrmions can be expected to scale linearly in time with diffusion constant given by:

\be
\langle |\mathbf{x}|^2\rangle (t)=2k_BT\frac{\alpha D}{G^2+(\alpha D)^2}t.\label{eqn:diffusivity}
\ee

The remaining term, which we now proceed to discuss in detail, seeks to model all repulsive interskyrmion interactions and boundary effects:

\begin{equation}
\label{eqn:extforce}
\mathbf{F}_{\mathrm{Ext}} = \mathbf{F}_S+\mathbf{F}_{\mathrm{B}},
\end{equation}
where $\mathbf{F}_S$ and $\mathbf{F}_{\mathrm{B}}$ are, respectively, skyrmion-skyrmion and boundary repulsion terms.

As described in Appendix A, we will mainly consider magnetic materials with a saturation magnetization of $M_S=1400\,k\mathrm{A/m}$, exchange stiffness $A_{ex}=27.5\,p\mathrm{J/m}$, interface-induced DMI constant $D=2.05\,m\mathrm{J}/\mathrm{m}^2$, perpendicular magnetic anisotropy constant $K_u=1.45\,M\mathrm{J}/\mathrm{m}^3$, and $\alpha=0.1$. The choice of these values is consistent with interfacially stabilized skyrmions on Pt/Co/MgO nanostructures~\cite{boulle2016room}.

\subsection{Skyrmion Interactions}

Short range repulsions between skyrmions due to exchange-dominated deformations to their respective spin textures have already been theoretically and numerically discussed in the literature in the absence of magnetic dipole effects~\cite{lin2013particle}. However, dense skyrmion populations are known to exhibit long range order leading to the formation of regular lattice structures~\cite{muhlbauer2009skyrmion,woo2016observation, heinze2011spontaneous} for which dipole interactions may play a relevant role. To capture the net sum of these effects on a single skyrmion's Thiele dynamics, we introduce a force term $F_{S}$ in (\ref{eqn:extforce}) which sums all the two-body interactions among skyrmions in a given ensemble (in the vein of~\cite{lin2013particle}): 
\begin{eqnarray}
\mathbf{F}_{S}&=&\sum_{j} \mathbf{F}_{S-S}(\mathbf{d}_{ij})\label{eq:addforce}\\
\mathbf{F}_{S-S}(\mathbf{d})&=&\exp\left[-\frac{a_1d^2+a_2d+a_3}{d+1}\right]\mathbf{\hat{d}}\label{eq:interparticle}
\end{eqnarray}
where $d_{ij}$ is the distance between particles $i,j$ and $\mathbf{\hat{d}}$ the normalized distance vector. The specific exponential form was chosen such that the repulsion behaves gaussian-like at short range and scales like a simple exponential at distances greater than the typical skyrmion diameter. 

Whereas the long range exponential scaling is justified in the thin-film limit where stray field energies can be modeled by a local shape anisotropy (see Appendix B), the short range Gaussian scaling was chosen to fit the micromagnetic behavior observed at small inter-skyrmion distances. In fact, for separation distances smaller than the skyrmion diameter, the notion of a skyrmion as a particle is expected to lose its meaning since the complexity of the magnetic texture voids the assumptions giving rise to the Thiele approximation. As such, even though we include a gaussian correction to the skyrmion repulsion, we will not consider scenarios where this correction is relevant. This typically happens whenever the interparticle spacing is smaller than the average skyrmion diameter.

To fit the phenomenological parameters $a_k$ (see Methods for values), we performed micromagnetic simulations far from boundaries where the net two-body skyrmion-skyrmion interaction in the presence of dipole effects can be isolated. Two skyrmions were initially set at a very close distance from each other and allowed to relax over time (see Methods for details). 

\begin{figure}[!htbp]
	\centerline{\includegraphics[width=4in]{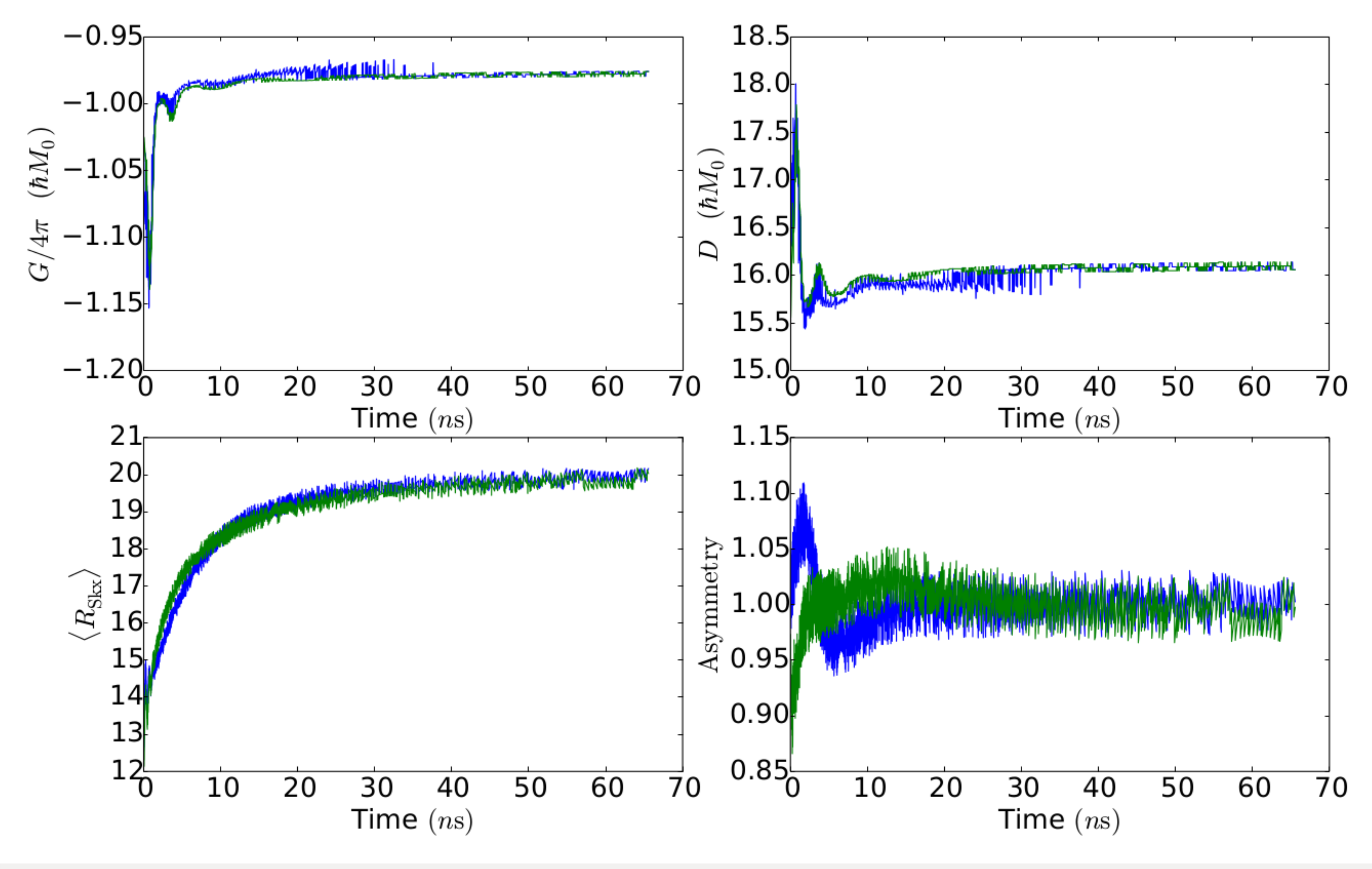}}
	\caption{{\footnotesize Two skyrmions, initially separated by $50\,n\mathrm{m}$, are allowed to evolve, via micromagnetic simulation, for $65\,n\mathrm{s}$ in the absence of thermal noise and applied currents. The topological constants are shown as a function of time. After a very brief relaxation phase, both the gyrovector $G$ (left) and the dissipative dyadic $D$ (right) stabilize around fixed values. Numerical deviations from constancy are due to the limited magnetic texture that was considered when performing integrals (\ref{eq:G} and \ref{eq:D}). To better qualify the relaxation process, the average skyrmion radius and skyrmion asymmetry (a value of $1.0$ representing perfectly circular particles) are plotted as a function of time.}}
	\label{fig:TwoSkx_GD}
\end{figure}

\begin{figure}[!htbp]
	\centerline{\includegraphics[width=4in]{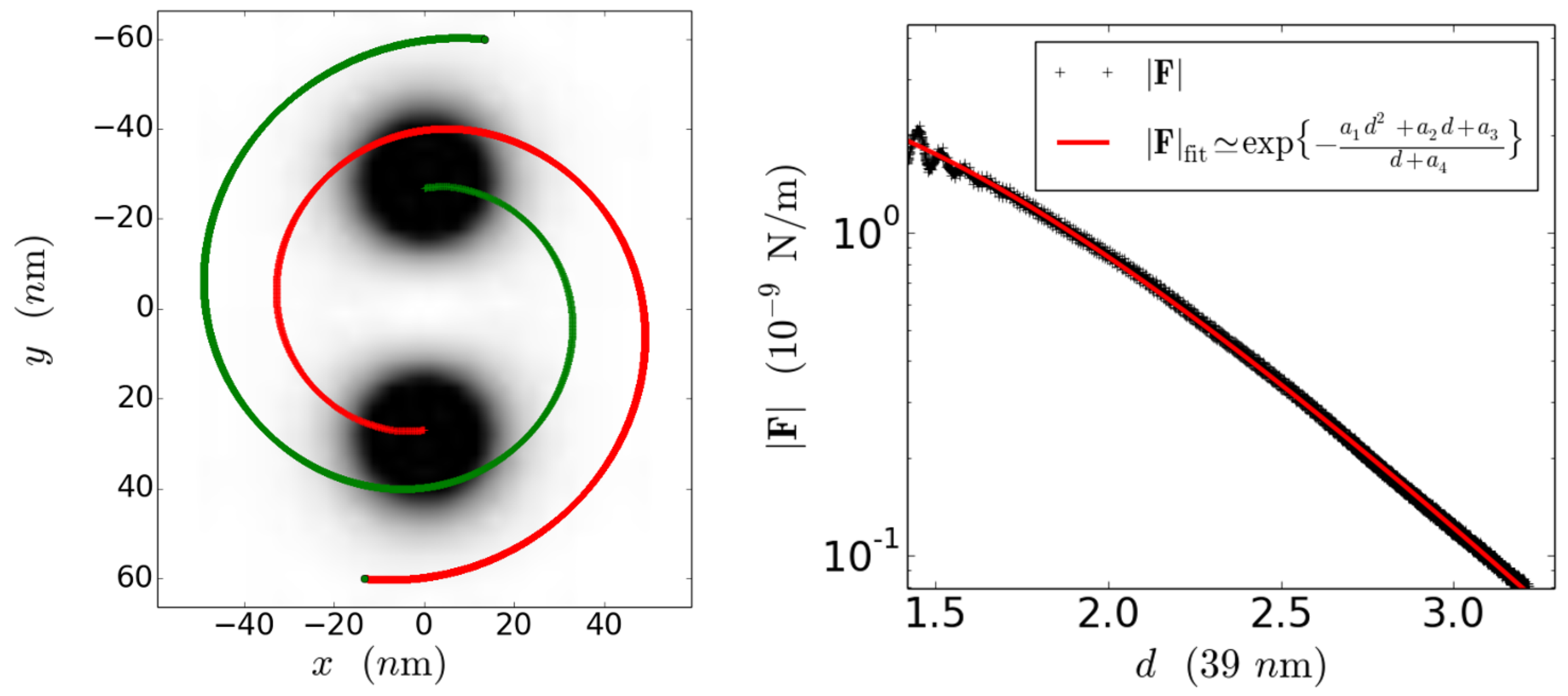}}
	\caption{{\footnotesize Two skyrmions, initially separated $50\,n\mathrm{m}$, are allowed to evolve for $65\,n\mathrm{s}$ in the absence of thermal noise and applied currents. Their extracted trajectories (left) are employed to compute the inter-skyrmion repulsion force intensity (right), plotted as a function of distance in units of average skyrmion diameter ($\sim 39\,n\mathrm{m}$), which was then fit phenomenologically according to equation (\ref{eq:interparticle}). Over the course of the simulation, the two skyrmions are seen to spiral away from each other as a direct consequence of the gyrotropic terms appearing in the Thiele-dynamical model (an initial frame of the simulation showing the skyrmion initial position is given as reference).}}
	\label{fig:TwoSkx}
\end{figure}

In Figure~\ref{fig:TwoSkx_GD}, we plot the gyrotropic matrix elements obtained by using (\ref{eq:G}-\ref{eq:D}) from the evolving magnetic profiles. Both $G$ and $D$ are seen to quickly stabilize onto equilibrium values, with $G=-4\pi$ as expected from the skyrmion's topological charge in the continuum limit~\cite{rajaraman1982solitons}. Employing these values, the net repulsion force between the two particles was extracted by computing $\mathbf{\hat{G}}\cdot\mathbf{\dot{x}}$ (the left hand side of (\ref{eqn:Evo})) as the simulation progressed. This in turn allows for the exploration of how the net repulsion force scales with the inter-skyrmion distance. Figure \ref{fig:TwoSkx} shows the particle trajectories (left) along with the extracted repulsion force as a function of inter-particle distance in units of skyrmion diameter (right). We find that the fit (\ref{eq:interparticle}) is adequate in modeling the observed two-body repulsion resulting from exchange and dipolar interactions. Particularly, we note that the Gaussian corrections to (\ref{eq:interparticle}) are only relevant at interparticle distances smaller than the average skyrmion diameter.

\subsection{Boundary Effects}

Skyrmions are experimentally studied on finite geometries. As such, it is essential to account for how the particle dynamics may be affected by the geometry's boundaries. The radial size of isolated skyrmions in magnetic dots is known to be confined by an explicit condition on the orientation of magnetic moments at the boundary~\cite{rohart2013skyrmion, sampaio2013nucleation}. This effect, a direct result of DMI in the material, is also responsible for repelling skyrmions from boundaries thus guaranteeing the transport properties which make them so useful for applications~\cite{sampaio2013nucleation,jiang2016direct}. As skyrmions in rarefied ensembles (as opposed to lattices) are capable of moving about freely, their dynamics eventually lead them close enough to the sample boundary where their behavior must be quantified through the Thiele formalism. 

\begin{figure}[!htbp]
	\centerline{\includegraphics[width=4in]{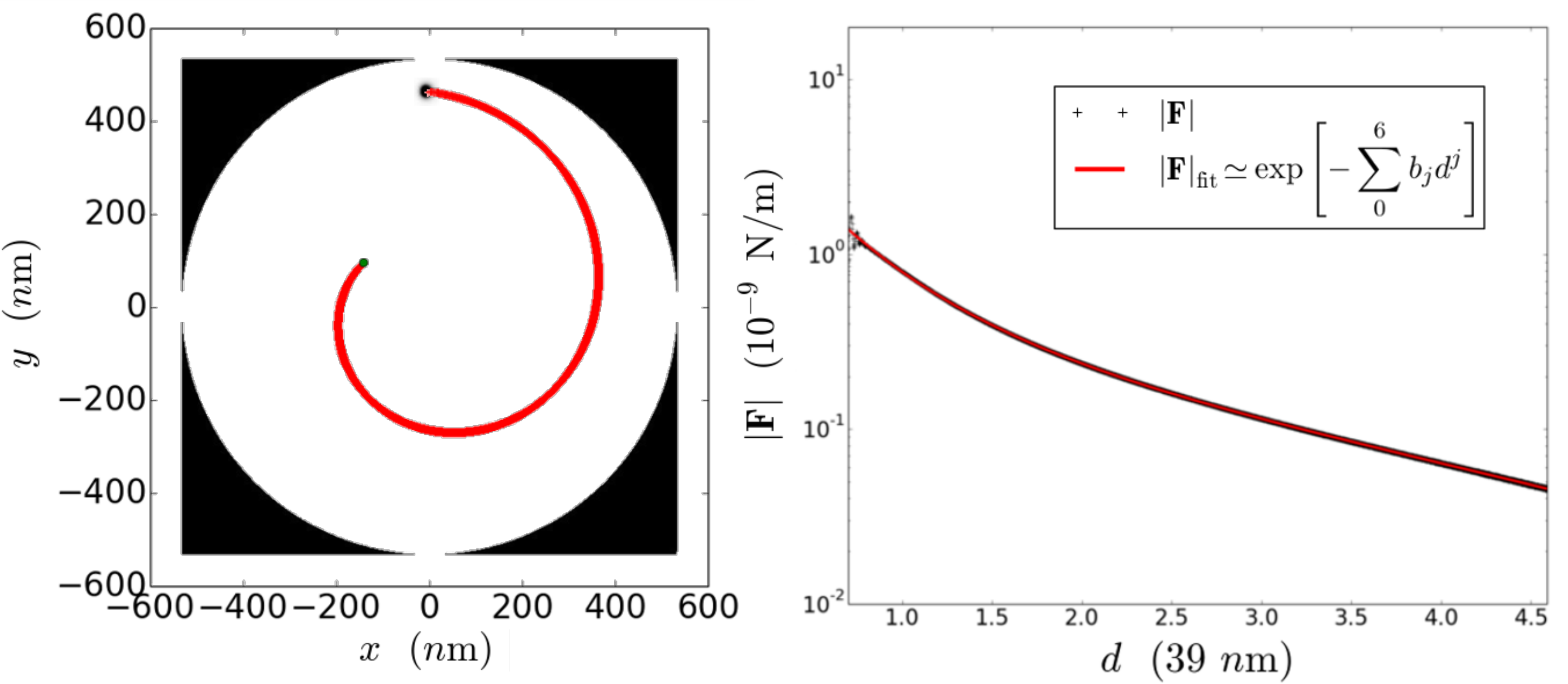}}
	\caption{{\footnotesize Net force experienced by a solitary skyrmion initially nucleated at a distance of $d=50\,n\mathrm{m}$ from the boundary of a $1024\,n\mathrm{m}$-diameter circular geometry and allowed to evolve for $1\,\mu\mathrm{s}$. Over the course of the simulation, the skyrmion's trajectory (left) is seen to move - clockwise - both along and away from the magnetic boundary (thick black curve) as a result of the gyrotropic effects captured by the Thiele-dynamical model (an initial frame of the simulation showing the skyrmion initial position is given as reference). The red line (right) shows a fit to the numerically derived force as a function of the distance from the boundary in units of average skyrmion diameter.}}
	\label{fig:Boundrep}
\end{figure}

We repeat our previous phenomenological analysis to model boundary effects by extracting the net force experienced by a solitary skyrmion initially placed adjacent to the boundary of a circular geometry. In Figure \ref{fig:Boundrep}, the force experienced by the skyrmion results in a net drift both along and away from the dot's boundary (left) whose scaling behavior as a function of distance from the boundary is distinctly different from that observed for skyrmion-skyrmion repulsions. In order to have an explicit phenomenological expression for the boundary repulsion force, we fit the observed data to a sextic gaussian form (see Methods):

\begin{equation}
\label{eqn:boundary}
|\mathbf{F}|_{\mathrm{B}}\simeq\exp\left[-\sum_0^6b_jd^j\right],
\end{equation} 
which is found to model the boundary interaction very well. 

Both the skyrmion-skyrmion and skyrmion-boundary interactions should not be considered valid when distances become smaller than the skyrmion diameter. At such length scales the topological texture of each skyrmion is radically altered leading to a behavior both not consistent with the Thiele dynamical model and with the notion of a topologically protected particles. When skyrmions are moved too close to each other or boundaries, annihilation typically ensues which, if anything, should be modelled via an attractive (rather than repulsive) force law~\cite{yoo2017current}.

\subsection{Ensemble Dynamics}

\begin{figure}[!htbp]
	\centerline{\includegraphics[width=5in]{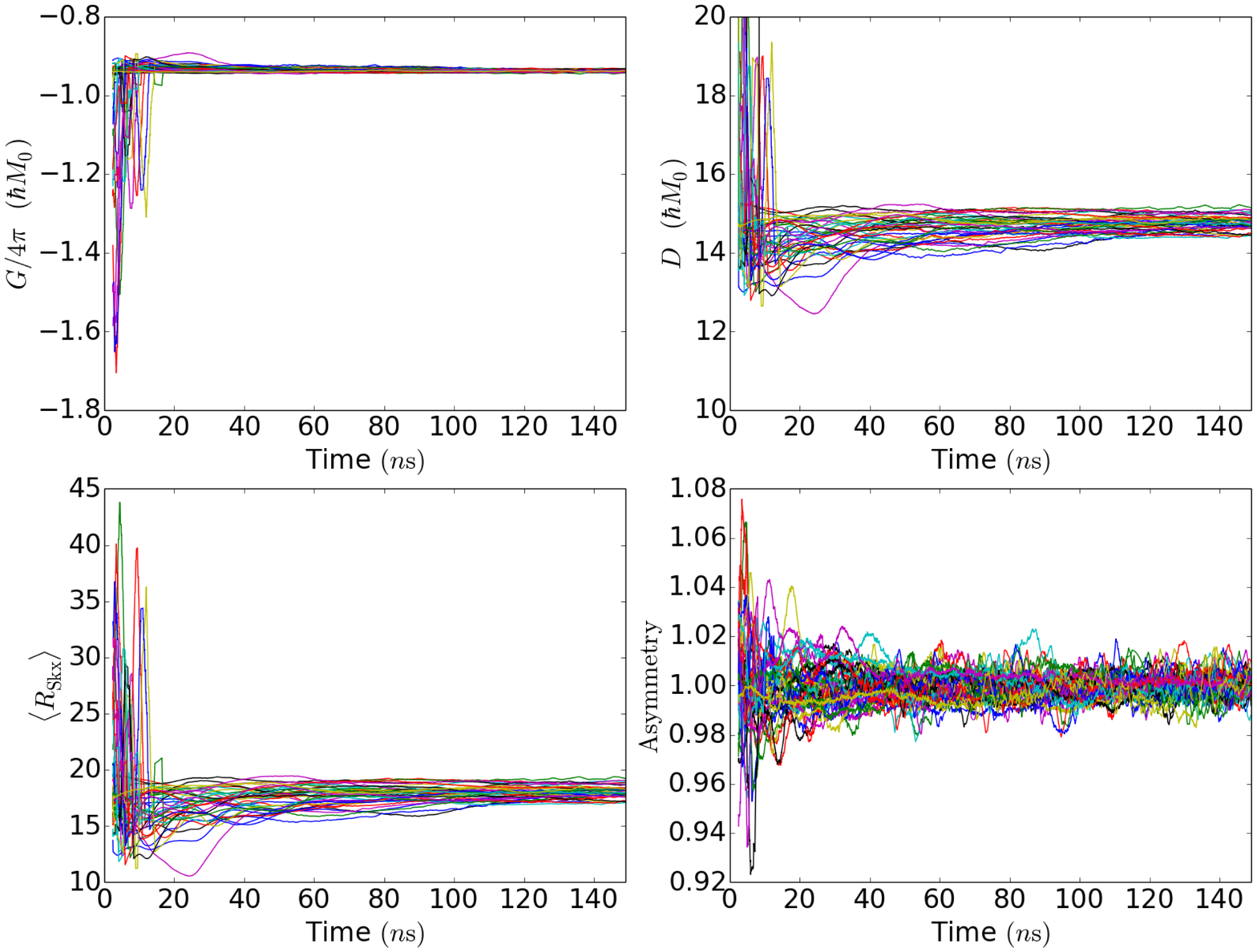}}
	\caption{{\footnotesize 47 skyrmions, initially placed randomly inside a $1024\;n\mathrm{m}$-diameter circular geometry, are allowed to evolve for $150\,n\mathrm{s}$ in the absence of thermal noise and applied currents. The topological constants are shown as a function of time. After a very brief relaxation phase, both the gyrovector $G$ (left) and the dissipative dyadic $D$ (right) stabilize around fixed values. Numerical deviations from constancy are due to the limited magnetic texture that was considered when performing integrals (\ref{eq:G} and \ref{eq:D}).}}
	\label{fig:20Thiele}
\end{figure}

Having modeled both two-particle and boundary repulsions, we now proceed to verify the assumptions of our model. We repeat the same numerical procedures for larger skyrmion numbers to verify that effective forces perceived by single skyrmions can be reconduced to an {\it n}-body sum of individual two-particle repulsions plus a boundary interaction term. We simulate 47 skyrmions initially placed randomly inside a circular geometry and allowed them to evolve for $150\,n\mathrm{s}$. In Figure~\ref{fig:20Thiele} we again see how the gyrotropic matrix elements of each skyrmion are consistent (within a $10\%$ margin of error) with those found for the two-particle simulation in Figure~\ref{fig:TwoSkx_GD}. In our particle treatment of the ensemble dynamics, we use identical Thiele parameters for every skyrmion as justified by the small error of our micromagnetically derived gyrotropic constants.

\begin{figure}[!htbp]
	\centerline{\includegraphics[width=5in]{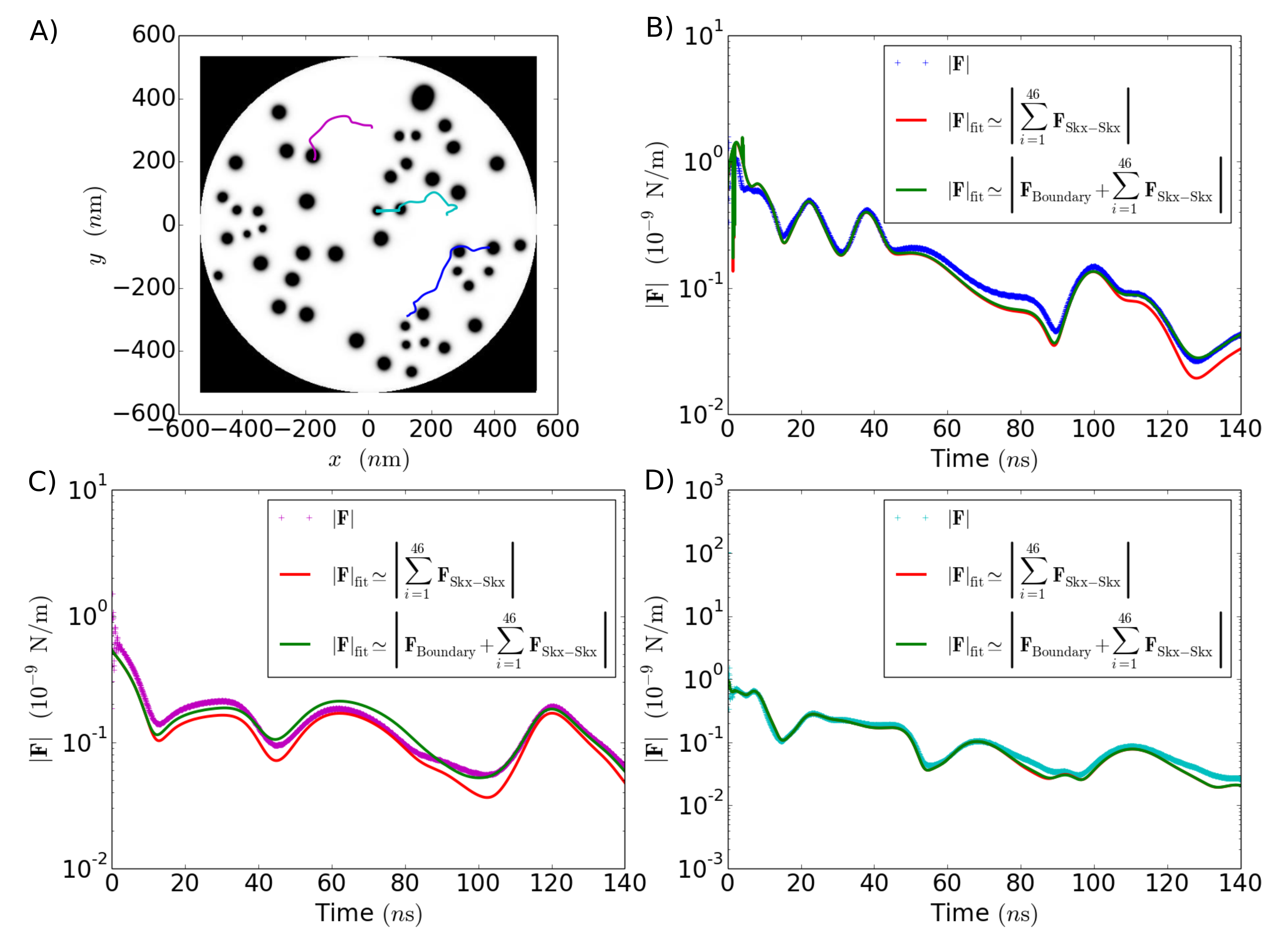}}
	\caption{{\footnotesize 47 skyrmions, initially placed randomly inside a $1024\;n\mathrm{m}$-diameter circular geometry are allowed to evolve for $150n\mathrm{s}$ in the absence of thermal noise and applied currents. Their extracted trajectories (a) were then used (an initial frame of the simulation showing the skyrmion initial positions is given as reference), in conjunction with their extracted gyrotropic constants (see Figure \ref{fig:20Thiele} and eq. (\ref{eqn:Evo})), to compute the effective forces (b-d) acting on each particle ($\mathbf{F}_{\mathrm{Ext}}=\hat{\mathbf{G}}^{\alpha}\cdot\mathbf{\dot{x}}$). The net forces acting on each sampled skyrmion is shown (colored crosses) and compared to the additive two-skyrmion prediction (\ref{eq:addforce}) both with and without the boundary interaction term (green and red fits respectively) to exemplify the importance of all the effects discussed in the text.}}
	\label{25Force_NOb}
\end{figure}

As discussed for the two- and one-particle studies of the previous sections, we reconstruct the net force experienced by each skyrmion from its trajectory and the gyrotropic parameters obtained from the two-skyrmion simulation. In Figure \ref{25Force_NOb}, the dynamics of three sample skyrmions are tracked and their net force extracted from the simulated dynamics.  We compare this to our phenomenological expressions accounting for just the skyrmion-skyrmion interactions (\ref{eq:addforce}) (red curves) and the case where boundary interactions are also included (\ref{eqn:boundary}) (green curves). The additional boundary term $\mathbf{F}_{\mathrm{B}}$ is included to highlight its importance in modeling skyrmions in the vicinity of the boundary, resulting in a much better fit. Particularly, the boundary effects only become important when skyrmions approach the boundary. As an example, in Figure \ref{25Force_NOb}-d, a particle whose trajectory stays close to the center of the nanodot is shown. In this circumstance, the particle's force fit with and without the boundary correction term are found to be virtually identical. The excellent agreement between micromagnetic simulations and our fit  demonstrates that our multi-body Thiele model accurately captures dipolar and exchange skyrmion-skyrmion interactions as well as skyrmion-boundary interactions. We now have all the tools necessary to efficiently model large ensembles of skyrmions in a device setting.

%% file: Sections/reshuffler.tex
Having evaluated the behavior of single skyrmions in a confined system and  extracted the forces and topological parameters of interacting skyrmions as well as the influence of sample boudaries, the next step towards the implementation of bio-inspired skyrmion-based devices is to apply the numerical techniques developed to simulate an actual real-world device. We begin with the Skyrmion Reshuffler introduced in Section 2 and then propose a completely different application of the same device as a neuron in the next section. Many other future devices employing the same principles as ours should be possible, as well as being numerically tractable via the modelling techniques just introduced. 

\begin{figure}
	\centerline{\includegraphics[width=6in]{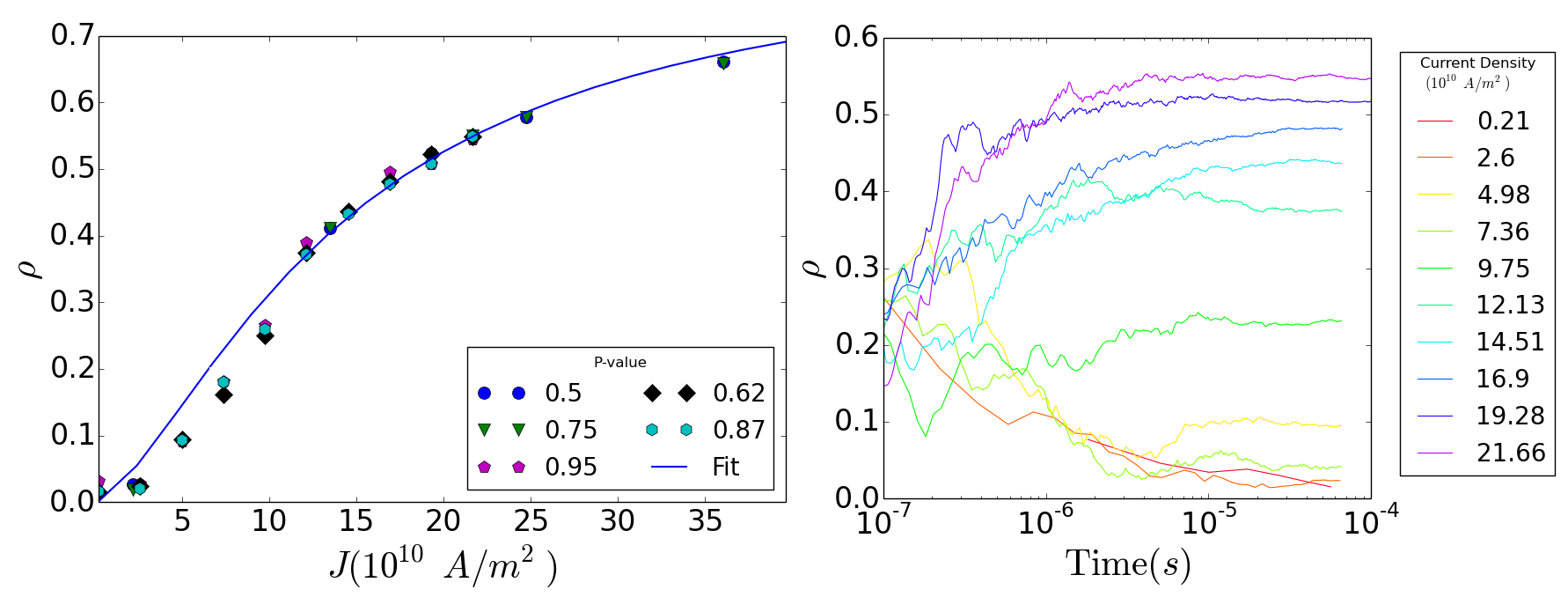}}
	\caption{{\footnotesize Simulation of the Skyrmion Reshuffler operating at a temperature of $300\,K$ employing two identical chambers ($1024\,n\mathrm{m}$ diameter) to scramble $0.5\,m\mathrm{s}$-long, randomly generated, telegraph noise input streams for different current intensities as measured at the $100\,n\mathrm{m}$-wide conduits. (Right) temporal evolution of correlation as the output stream is generated (data corresponds to the $p=0.62$ simulation). (Left) final product-moment correlation coefficient at each current intensity for five different input p-values. The blue curve fits data to the scaling $\rho= c_A/(c_B+J^{-3/2})$ argued in the discussion following (\ref{corrscale}).}}
	\label{ReshufflerSim}
\end{figure}

We simulate the Skyrmion Reshuffler in Figure 3 by considering two identical circular $1024\,n\mathrm{m}$ chambers into which we inject skyrmions every $5\,n\mathrm{s}$ depending on the state of a $0.5\,m\mathrm{s}$-long telegraph noise signal with fixed p-value. The skyrmions are initially injected from $100\,n\mathrm{m}$-wide input conduits with currents of varying intensity (see Methods). After drifting and diffusing through their respective chambers, they are then {\it read} upon arriving at the output track and used to reconstruct an outgoing signal. The output signal's correlation to the input is then checked by computing the product-moment correlation coefficient of the two~\cite{fisher1915frequency}:

\begin{equation}
\rho_{X,Y}=\frac{\mathrm{cov}(X,Y)}{\sigma_X\sigma_Y},
\end{equation}
where $\mathrm{cov}(X,Y)$ is the covariance of two signals $X$ and $Y$ while $\sigma$ is their respective variance. We show this result in Figure \ref{ReshufflerSim} as a function of the drift current intensity for different signal $p$-values. 

\begin{figure}
	\centerline{\includegraphics[width=2.5in]{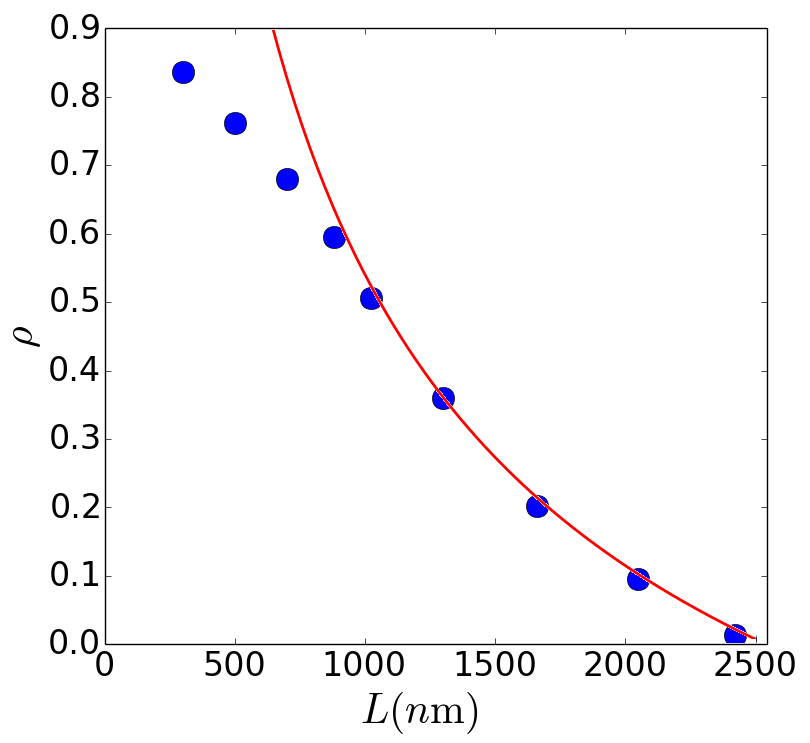}}
	\caption{{\footnotesize Output signal correlation as a function of different chamber sizes driven at a current intensity of $1.7\cdot 10^{11}\;A/\mathrm{m}^2$. Increasing the chamber size results in a longer diffusion time of the particles inside the chamber leading to a better decorrelation of the output signal. The red curve fits data to the scaling $\rho=c_A+c_B/\sqrt{L}$ argued in the discussion following (\ref{corrscale}).}}
	\label{multichambers}
\end{figure}

If the chamber is significantly larger than the transverse size of the input/output conduit tracks, current densities inside the chamber may become small enough, allowing skyrmions to interact and diffuse thermally in the chambers before exiting. In Figure~\ref{ReshufflerSim}-left, the correlation coefficient $\rho$ is in fact seen to decrease rapidly with lower current intensities. For the physical values considered, our device shows that strong decorrelation is achieved for currents intensities of $\sim 10^{10}\;\mathrm{A}/\mathrm{m}^2$~\footnote{For the lowest currents sampled no more than $51$ particles were observed in either chamber at any given time. As we will see later on (see Figure~\ref{maxdensity}), this number is well below the saturation limit of $90$ for this chamber size.}. From the moment the output signal reconstruction commences, the correlation reaches its steady state value on a $\mu\mathrm{s}$ timescale as can be seen in Fig.~\ref{ReshufflerSim}-right by the flattening of the correlations' temporal traces. The time required to inject and read-out a statistically significant number of skyrmions determines the precise timescale for relaxation of correlations towards their steady-state equilibrium. This leads to the intuitive understanding that stronger currents allow the device to function faster at the cost of decorrelation efficiency. In turn, what exactly qualifies as 'satisfactory decorrelation' will ultimately depend on the details and specifications of the stochastic computing circuit one wishes to implement. Furthermore, we don't see significant deviation in the results as a function of the p-value assigned to the input signal. In Fig.~\ref{multichambers} we instead fix the current intensity at $1.7\cdot 10^{11}\;A/\mathrm{m}^2$ and allow for the chamber sizes to vary. Larger chambers allow the particles to diffuse for longer thus resulting in a better decorrelation of the shuffler's output signal. Ultimately, Figures~\ref{ReshufflerSim} and~\ref{multichambers} show how the degree of decorrelation achieved can be tuned by both lowering the current intensity and/or increasing the chamber sizes.

Ideally, one would like the signal scrambler to work as fast as possible. The speed at which a statistically significant output signal can be generated depends directly on the intensity of the driving current moving the injected skyrmions through the chambers (we neglect effects due to finite response times of nucleation and read-out mechanisms). However, as shown in Figure \ref{ReshufflerSim}, higher currents lead to highly correlated output signals due to the fact that skyrmion particles do not spend enough time diffusing and interacting in the chambers. In Fig.~\ref{ReshufflerSim}-left we find $64\%$ correlation at the highest driving current sampled ($3.8\cdot10^{11}\;\mathrm{A}/\mathrm{m}^2$).

Alternatively, one may try to increase the temperature in an attempt to amplify thermal diffusion. This, however, leads to both a higher energetic consumption as well as the potential for an increased skyrmion annihilation rate. Alternatively, one could attempt to leverage the presence of random material inhomogeneities to further drive the diffusive processes. This has been recently discussed and characterized in the literature~\cite{kim2017current} as a possible technique to tune the skyrmion dynamical properties. Ultimately, the condition affecting this trade-off lies in the relative intensity of dynamical drift and thermal noise. From (\ref{eqn:diffusivity}) one has that the skyrmion's diffusivity over time results in a dispersion given by:

\begin{equation}
\langle|\mathbf{x}|^2\rangle(t)=2\frac{k_BT}{\gamma M}\left(\frac{\alpha D}{G^2+(\alpha D)^2}\right)t,
\end{equation}
where $t$ is the elapsed time and angle brackets $\langle\cdot\rangle$ refer to averaging over noise. 

Significant decorrelation between input and output signals takes place whenever this diffusion is large enough to allow scrambling in the sequence of skyrmions being injected into the chamber. Denote then by $l_0$ the relative spacing between successively injected skyrmions. If thermal effects manage to displace skyrmions more than their inter-spacing ($\langle|\mathbf{x}|\rangle \gg l_0$) over the time it takes them to cross the chamber, it is reasonable to expect the ingoing and outgoing sequences to become randomized relative to each other. As such, a qualitative estimate of the correlation between input and outgoing signals can be found in the ratio between these two length scales $\rho\propto 1/(1+\sqrt{\langle|\mathbf{x}|^2\rangle}/l_0)$. Recalling then from (\ref{eqn:Evo}) that the particle drift speed scales with the current intensity ($|\dot{\mathbf{x}}|\simeq\mathbf{v}_s\propto J$) one can expect that correlations will scale relative to system parameters as:

\begin{equation}
\label{corrscale}
\frac{\rho}{1-\rho}\propto\frac{l_0}{\sqrt{\langle|\mathbf{x}|^2\rangle(t_{\mathrm{exit}})}}\propto \nu\sqrt{\frac{\gamma M}{k_B}\left(\frac{G^2+\alpha D^2}{\alpha D}\right)\frac{J^3}{LT}},
\end{equation}
where, denoting the chamber diameter as $L$ and the injection frequency as $\nu$, the time taken to cross the chamber is $t_{\mathrm{exit}}\simeq L/|\mathbf{v_s}|\propto L/J$ and the inter particle spacing is $l_0\propto \nu\,J$. The fitting curves in Figures~\ref{ReshufflerSim} and~\ref{multichambers} indeed verify that the scalings $\rho/(1-\rho)\propto J^{3/2}/\sqrt{L}$ match our numerical data, thus justifying how correlations will be lower for smaller currents and larger chamber sizes.

A potential improvement to the Skyrmion Reshuffler can theoretically be found in the use of anti-ferromagnetic (AFM) materials where the possibility of stabilizing AFM skyrmions has recently been suggested~\cite{barker2016static, zhang2016antiferromagnetic,jin2016dynamics}. At the core, the main differences in the material parameters between AFM and FM devices lies in the sign of the exchange interaction and the magnitude of the magnetostatic interaction. These differences have potentially large repercussions on the Thiele modeling of skyrmions in the AFM systems. The negative exchange interaction leads to a vanishing gyrovector $G= 0$, implying two important consequences. First, the AFM skyrmion always has zero transverse velocity relative to the direction of current flow. This, in FM skyrmions, is true exclusively in the singular scenario $\alpha=\eta$. Secondly, and most importantly, a vanishing gyrovector implies a larger diffusion constant (\ref{eqn:diffusivity}) and, consequently, smaller correlations (\ref{corrscale}). Depending on the value of the damping constant $\alpha$ in the materials employed ($\alpha\sim 0.01-1$), the difference in magnitude of the diffusion constant in FM and AFM skyrmions can easily span an order of magnitude difference. The magnitude of the magnetostatic interaction also complements the AFM skyrmion as it may lead to the stabilization of much smaller, and more thermally sensitive, particles. For all these reasons, one expects that AFM skyrmions would allow operation of our proposed device at higher currents and faster speeds.

Any accidental particle nucleation or annihilation taking place in the chambers will not influence the outgoing signal's p-value as long as its rate remains proportional to the total number of skyrmions present in each chamber. This proportionality is guaranteed for accidental particle annihilations as it will depend on how many skyrmions are present in the chamber at any given time. Nucleations on the other hand can be potentially problematic for the device's proper functioning. Luckily, the experimental observations of skyrmions at room temperature show no signs of thermally influenced skyrmion instabilities. This leaves the interactions with material defects and background inhomogeneities~\cite{bacani2016measure} as the only other potential source of skyrmion anihilations. As long as both chambers are similarly constructed, however, this will lead to a similar rate of annihilations in both chambers thus preserving input-to-output p-values modulo a possible decrease in the output signal's frequency.

Overall, given the DC-current driving the device (and possibly nucleating the necessary skyrmions as in~\cite{everschor2016skyrmion}), we expect the skyrmion reshuffler to operate at very low energies ($\sim\mu\mathrm{W}$). A more accurate metric would involve calculating the energy cost per reshuffled {\it bit} of information being transmitted through the telegraph signal. Typical CMOS-based solutions reshuffle clocked stochastic bitstreams~\cite{tehrani2010relaxation} whose energy consumption can be estimated at $\sim 10\;p\mathrm{J}/\mathrm{bit}$. Placing their large area imprint problems aside, they are expected to perform efficiently only for slow bitstreams as their energy consumption will scale linearly with the signal's frequency. On the other hand, our Skyrmion Reshuffler constant energy costs will prove ideal for reshuffling high frequency streams instead. As already mentioned, the device modeled in this section considered a constant skyrmion injection rate of $0.2\;\mathrm{nucleations}/n\mathrm{s}$ thus allowing us to potentially reshuffle signals with frequencies as high as $5\;G\mathrm{Hz}$, resulting in an equivalent energy-per-bit cost of $5\;f\mathrm{J}/\mathrm{bit}$ which is orders of magnitude lower. 

%% file: Sections/neuron.tex
\begin{figure}
	\centerline{\includegraphics[width=3in]{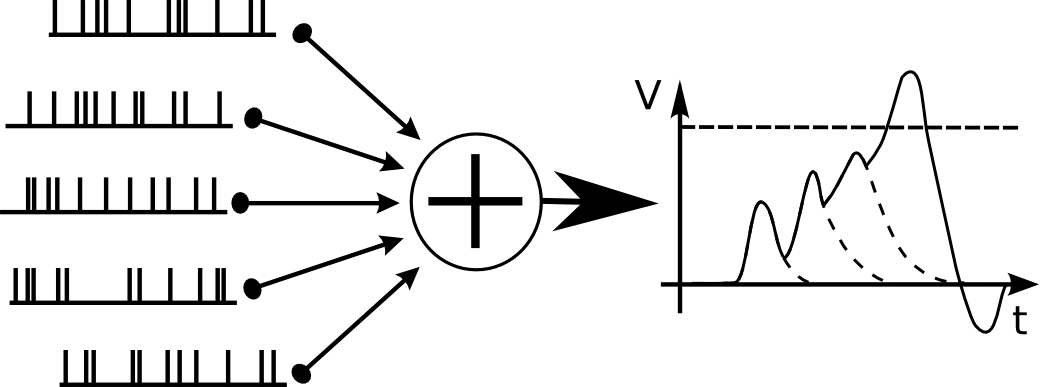}}
	\caption{{\footnotesize A neuron collects neuronal spiking activity from nearby neurons. The incoming spikes lead to an increasing voltage potential between the inside and outside of the receiving neuron. Once the potential difference reaches a certain threshold, the receiving neuron discharges by releasing a voltage spike to its axon terminals where the signal is transmitted through synapses to other neurons. Due to voltage leakage across the cell membrane, the neuron can be thought as a leaky capacitor with low breakdown voltage.}}
	\label{Neuron_intro}
\end{figure}

We now turn to demonstrate how a similar design can be used to emulate the behavior of a neuron. Neural functioning relies on the ability to transmit and receive electrochemical signals to/from a large number of interconnected neurons by means of its {\it dendritic} connections (see Fig.~\ref{Neuron_intro}). As coupled neurons spike, the receiving neuron registers the event with a voltage increase (or decrease) across its cellular membrane. The underlying physical mechanism for this behavior is the presence of ion channels scattered over the cell's membrane responsible for these potential differences. Such physical channels are not ideal however and {\it leak} over time resulting in a steady decay of the neuron's voltage towards its resting potential in the absence of incoming electrochemical pulses. As a result of this behavior, a neuron can be thought of as a leaky capacitor. 

If incoming neuronal activity is registered at a high enough frequency, however, the receiving neuron overcomes its leaking voltage and steadily charge. A threshold potential exists beyond which the neuron suddenly discharges and resets. This basic model of neuronal functioning is known as the {\it leaky integrate-and-fire} neuron model~\cite{hodgkin1952quantitative,koch1998methods}. The neuron effectively serves as a basic memory integrating the sum of past voltage spikes received with the added property that this memory is lost over time if a potential threshold is not reached. 

The analogies to our skyrmion reshuffler are already apparent as the injection of particles into it can be thought of as a response to a voltage spike train (shown in Fig. 3). In this sense, the collecting of skyrmions inside the chambers serves as a history of the input spike train. If the passage of skyrmions through the output conduit were to be precluded (such as with a voltage gate~\cite{kang2016voltage,li2017magnetic}), skyrmions progressively accumulate inside the geometry where their density can be estimated via a reading element placed inside each chamber (see Fig.~\ref{Neuron}). Each chamber can then be thought of as a reservoir collecting the memory of whatever input signal was employed to populate it. 

\begin{figure}{r}
	\centerline{\includegraphics[width=3in]{Figures/Fig_Neuron.pdf}}
	\caption{{\footnotesize The Skyrmion Neuron consists of a magnetic chamber into which skyrmions are injected depending on the state of one or multiple input telegraph noise signals. A voltage gate, initially in an ON state, will impede the skyrmions from drifting to the output conduit thus leading to particle accumulation. A reading element placed inside the chamber can be used to estimate the skyrmion density inside the chamber and switch off the voltage gate when a critical density has been reached.}}
	\label{Neuron}
\end{figure}

Furthermore, an upper limit exists to how many skyrmions may be crammed into each such chamber. In fact, as skyrmions become compressed with their neighbors to length scales comparable to the typical skyrmion diameter, the combination of dipolar strain among particles along with thermal effects breaking topological stability may cause them to annihilate. The topological stability of the skyrmion profile is only guaranteed under continuous deformations of the magnetic texture~\cite{bogdanov2001chiral,rossler2006spontaneous}. The introduction of a sufficiently large thermal field in the magnetization dynamics can make a skyrmion collapse with some probability~\cite{romming2013writing, koshibae2014creation, hagemeister2015stability, siemens2016minimal, oike2016interplay, yin2016topological}. Alternatively, and possibly more importantly, skyrmion annihilation can be biased by the presence of randomly scattered material defects in the magnetic sample populated by the skyrmions. As such, small room temperature skyrmions can potentially annihilate (and nucleate) simply due to thermal fluctuations and material disorders. The maximal particle density inside a chamber can hence be thought of as a saturation threshold beyond which memory of the input signal will certainly be destroyed. 

\begin{figure}
	\centerline{\includegraphics[width=2.5in]{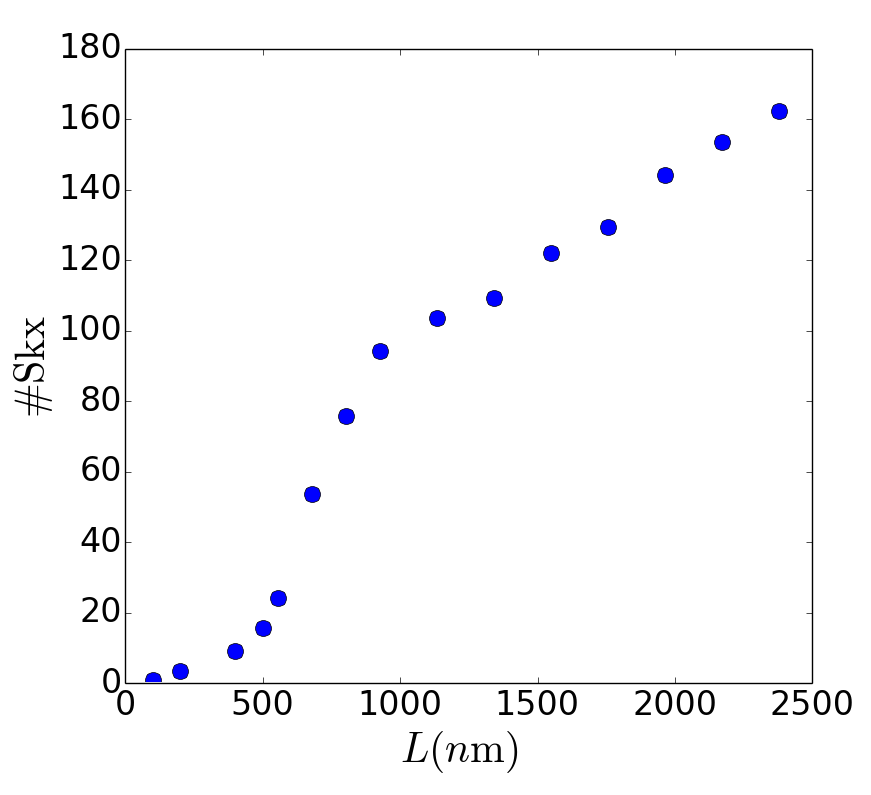}}
	\caption{{\footnotesize Maximum number of $\sim 39\, n\mathrm{m}$ skyrmions that can be stabilized in a circular chamber as a function of chamber diameter at $300\,K$.}}
	\label{maxdensity}
\end{figure}

In Fig.~\ref{maxdensity}, we perform micromagnetic simulations at $300\,K$ to establish the maximal number of $\sim 39\, n\mathrm{m}$ skyrmions which can be stabilized inside chambers of varying sizes. We nucleate the skyrmions randomly in the chamber in ever increasing numbers and allow the magnetic texture to stabilize for $2\,n\mathrm{s}$ before counting how many skyrmions are left. This process was then repeated ten times before computing the average. For each chamber diameter explored, a saturation value is found beyond which no more skyrmions can be added reliably into the geometry.

The combination of memory due to particle conservation and memory-loss from annihilations allows one to view our proposed skyrmionic chamber as a lossy memory storage device which integrates an incoming signal while allowing a certain decay of the accumulated information. Upon switching off the voltage gate, the collected skyrmions can be released into the output conduit whenever a certain threshold skyrmion density is reached (as measured by a reading element inside the chamber). Following these principles, the functioning of our device behaves as a leaky integrate-and-fire neuron where, instead of integrating discrete current pulses through electric potential accumulation before discharging, our device collects skyrmions instead and successively dumps them into the output conduit.

%% file: Sections/conclusion.tex
The impracticality of performing complex micromagnetic simulations out to long timescales currently poses an obstacle to the design and proof-of-concept of novel spintronic-based devices with a $\sim\mu\mathrm{m}$ area imprint. Under suitable simplifying assumptions however, the behavior of solitonic magnetic spin-textures can be radically simplified and reduced to classical {\it n}-body problems which can be tackled through a plethora of numerical techniques. We have shown how one such {\it particle} model (the {\it Thiele} equation) can be phenomenologically tuned in such a way to account for complex exchange and dipole mediated interactions among stable magnetic skyrmions. By means of the tools developed, we proposed and modeled two skyrmion devices whose behavior would be computationally intractable via standard micromagnetic techniques.

The Skyrmion Reshuffler encodes a telegraph noise signal into a sequence of skyrmions injected and current-driven through two separate chambers. The dominance of thermal diffusion in each chamber scrambles the particle order in such a way that the output sequence can be used to reconstruct an output signal with negligible correlations to the input one. Through simple scaling analyses, we have shown how the system parameters impact these correlations and balance the trade-offs between speed and effectiveness of the device's operation. Leveraging the interplay between skyrmion particle stability and thermal diffusivity, we have argued that the skyrmion reshuffler stands to solve a long standing problem in the field of stochastic computing where computations defined over stochastic signals require correlations to be absent for proper functioning. The spintronic basis of the skyrmion reshuffler's design allows for a highly attractive area imprint and energy costs.

The Skyrmion Neuron adapts the skyrmion reshuffler's design to propose a device capable of emulating the integrate-and-fire characteristics of a neuron. By means of an added voltage gate selectively prohibiting the outflow of skyrmions from the chamber, as well as an extra read-out element {\it inside} the chamber, the skyrmion density in the chamber can be measured and the voltage gate released upon a certain threshold being reached. In doing so, the chamber integrates the signal intensity (or length) driving the input sequence of skyrmions injected into the chamber. To complete the analogy with the animal neuron, we argue that skyrmions perform the role of both voltage accumulation as well as neurotransmitters in this specific layout. The devices we propose are compatible with standard CMOS and can also be the building block of future fully spintronic probabilistic computers.

Overall, the sensitivity of magnetic soliton structures to small driving currents and thermal noise makes them optimal candidates for the development of novel probabilistic devices. Furthermore, the {\it nano}-scale size of these particles implicitly guarantees that devices employing them will be both energy efficient and scalable. Our results suggest that the basic research currently being done on these exotic magnetic solitons fertilizes the field for future disruptive applications to come.

%% file: Sections/acknowledgements.tex
This work was funded by the European FET OPEN `{\it Bambi}' project 618024 and the European Research Council ERC grant `{\it bioSPINspired}' 682955. J.V.K and V.C acknowledge FET OPEN `{\it MAGicSky}' project 665095. The authors would like to thank A. Thiaville for his very useful comments during the revision of the manuscript.

%% file: Sections/appA-methods.tex
The three-dimensional micromagnetic simulations were performed using the MuMax3 GPU-accelerated micromagnetic simulation program~\cite{vansteenkiste2014design}. The average energy density contains the exchange energy, the anisotropy energy, the applied field (Zeeman) energy, the magnetostatic (demagnetization) energy and the DMI energy terms. In all simulations, the thickness of the magnetic nanotracks was $2\,n\mathrm{m}$. The length of the input/output conduits was $50\,n\mathrm{m}$, while the width was set to $100\,n\mathrm{m}$. Magnetic parameters used in the simulations: saturation magnetization $M_S=1200\,k\mathrm{A/m}$, exchange stiffness $A_{ex}=8\,p\mathrm{J/m}$, interface-induced DMI constant $D=2.1\,m\mathrm{J}/\mathrm{m}^2$, perpendicular magnetic anisotropy constant $K_u=0.9\,M\mathrm{J}/\mathrm{m}^3$ and gyromagnetic ratio $\gamma=-2.211\cdot 10^5\,\mathrm{m}/\mathrm{A}\mathrm{s}$. The Gilbert damping coefficient $\alpha$ was set to $0.1$. The choice of these values is consistent with interfacially stabilized skyrmions on Pt/Co/MgO nanostructures~\cite{boulle2016room}.

All models are discretized into tetragonal cells with the constant cell size of $2\times 2\times 2\,n\mathrm{m}^3$ in the simulations, whose linear size is smaller than both the fundamental length scale $A_{ex}/D\simeq 4\,n\mathrm{m}$ and the domain wall width $\sqrt{A_{ex}/K_u}\simeq 3\; n\mathrm{m}$. With these parameters, we observed the stabilization of skyrmions with an average of diameter of $\sim 39\, n\mathrm{m}$ both at zero and $300\,K$ temperature. The integration time steps used were $5\cdot 10^{-12}\,\mathrm{s}$ and $5\cdot 10^{-14}\,\mathrm{s}$ at $0\,K$ and $300\,K$ respectively. At finite temperature breathing modes were observed at GHz frequencies. 

The tracking of skyrmion particles was performed on dynamical snapshots taken at each time step using the python implementation of the DLIB open-source machine learning library~\cite{king2009dlib}. The classic Histogram of Oriented Gradients (HOG) feature combined with the linear classifier, image pyramid, and sliding window detection scheme were used to train a detector to recognize skyrmions using a training set of $200$ stabilized skyrmion profiles. From the geometry's total magnetization profile, the tracking scheme allowed us to obtain a set of regions containing each one skyrmion only. 

The computation of the skyrmion center was performed by first fitting the magnetic profile of each region with a cubic spline interpolation and then finding the maximum z-component of the magnetization on the interpolated profile. The algorithm allows for a $\sim p\mathrm{m}$ resolution of the skyrmion center. The gyrotropic parameters were computed by performing the relelvant surface integrals on the interpolated profile of each skyrmion particle as discussed in equations (\ref{eq:G}) and (\ref{eq:D}).

The {\it n}-body Thiele dynamics were solved using a specialized GPU-accelerated CUDA/C++ solver developed by the authors. The stochastic dynamical system of equations was solved employing a Heun scheme ensuring convergence to the proper Stratonovich solution. The current density distributions were first obtained by simulating current flow on COMSOL~\cite{multiphysics2012comsol} for the geometry discussed. Room temperature skyrmions are known to exhibit breathing modes~\cite{kim2014breathing}, these will however not impact the net dynamics due to the gyrotropic terms being effectively scale-free in the particle size~\cite{schutte2014inertia} when the skyrmion size is much larger than the domain wall width $\sqrt{A_{ex}/K_u}\simeq 4.35\; n\mathrm{m}$. Qualitatively different dynamical behavior can however be expected for large skyrmions where an extra mass term is expected to play a role in the Thiele-description. For the material parameters in question, the numerical fitting constants for the skyrmion-skyrmion two-body repulsions were found to be $\mathbf{a}\simeq[2.709,-5.643,0.964]$, whereas the fitting constants for boundary repulsions were $\mathbf{b}\simeq[0.001419, -0.02631, 0.24135, -1.1609, 3.3547,-2.1786]$.  In this work, we have taken $\eta=\alpha$ and $\mu=1$ for convenience.

%% file: Sections/appB-1Drep.tex
A magnetic texture $\mathbf{M}=\mathbf{M}(x,y,z)$ with constant magnetic saturation $|\mathbf{M}|=M_S$ extended over a 3-dimensional domain $\Omega$ of thickness $d$ can be described by a micromagnetic free energy which, in the presence of an out-of-plane uniaxial anisotropy (along the $z$-axis) and an interfacial Dzyaloshinskii-Moriya interaction ({\it DMI}), takes the form~\cite{bogdanov1994thermodynamically,thiaville2012dynamics}:

\bea
E_{\mathbf{M}}&=&\frac{A}{M_S^2}\int_{\Omega}\mathrm{d}^3r\;|\nabla\mathbf{M}|^2+\frac{K}{M_S^2}\int_{\Omega}\mathrm{d}^3r\;|M_z|^2\\
&+&\mu_0\int_{\Omega}\int_{\Omega}\mathrm{d}^3r\mathrm{d}^3r'\;\frac{\nabla\cdot\mathbf{M}(\mathbf{r})\nabla\cdot\mathbf{M}(\mathbf{r'})}{8\pi|\mathbf{r}-\mathbf{r'}|}\\
&+&\frac{Dd}{M_S^2}\int\int\mathrm{d}x\,\mathrm{d}y\;\mathbf{M}\cdot\left(\partial_x\mathbf{M}\times\partial_y\mathbf{M}\right),
\eea
where $A$, $K$ and $D$ are the exchange stiffness, anisotropy and Dzyaloshinskii-Moriya strength respectively. In order of appearance, the terms appearing in the magnetic free energy are: the exchange term, which prefers constant magnetization configurations; the magnetocrystalline anisotropy, which favors out-of-plane magnetization configurations; the magnetostatic term, which prefers divergence-free configurations; and the surface DMI term, which favors chiral symmetry breaking. 

In the limit of ultrathin ferromagnetic films where the sample thickness is smaller than the exchange length $l_{ex}$ ($d\leq l_{ex}= \sqrt{2A/(\mu_0M_S^2)}$) the magnetic free energy above can be described by a model where the stray field energy is expressed as a local shape anisotropy term~\cite{gioia1997micromagnetics,knupfer2017magnetic}. This model can in turn be used to treat the simplified 1-dimensional case of magnetic wires in the presence of DMI with magnetic energy given by~\cite{muratov2017domain}:

\be
E_{\mathbf{M}}=\int_{-\infty}^{\infty}\mathrm{d}x\;\left[|\partial_x\theta|^2+Q^2\sin^2\theta\right]-\kappa\pi,
\ee
where $\kappa=D\sqrt{\frac{2}{\mu_0M_S^2A}}$, $Q=\sqrt{\frac{2K}{\mu_0M_S^2}-1}$, and the magnetic profile has been parametrized as $\mathbf{M}=M_S\,(\sin\theta,0,\cos\theta)$. The dimensionless constant $Q$, known as the {\it quality factor}, captures both the contributions of the crystalline anisotropy and the stray field in the ultrathin film limit considered. The energy is minimized by domain wall solutions whose explicit form can be verified to be:

\be
\theta(x)=2\,\atan\left[e^{-Qx}\right].
\ee

Let us then consider a magnetic texture consisting of two such domain walls separated by a distance $\delta$:

\be
\theta(x)=\theta_+(x)+\theta_-(x)=2\,\atan\left[\eta e^{Qx}\right]+2\,\atan\left[\eta e^{-Qx}\right],
\ee 
with $\eta=\exp(-Q\delta/2)$. The total energy of the system can be written as:

\be
E_{\theta}=E_{\theta_+}+E_{\theta_-}+E^{\mathrm{int}}_{\theta_+,\theta_-}(\delta)
\ee
\be
 \begin{aligned}
 E^{\mathrm{int}}_{\theta_+,\theta_-}(\delta)=\int_{-\infty}^{\infty}\mathrm{d}x\;\left[\partial_x\theta_+\partial_x\theta_-+\frac{1}{2}\sin2\theta_+\sin2\theta_- \right.\enspace &
 \\
 \left. -2\sin^2\theta_+\sin^2\theta_-
 \right]&
 \end{aligned}
\ee
where the first two terms are simply the minimum energy of each domain wall taken independently while the third measures the interaction potential of the two walls. The repulsive force between the two domain walls will then be straightforwardly given by the derivative of interaction potential with respect to the distance separating the walls ($F_{\theta_+,\theta_-}(\delta)=-\partial_{\delta}E^{\mathrm{int}}_{\theta_+,\theta_-}$). Tedious, but fairly elemental analysis then results in the following repulsion force:

\bea
F_{\theta_+,\theta_-}(\delta)&=&2Q\eta^2(I_{1,1,1,0}-4I_{1,1,1,2})+2^5\eta^4(2I_{2,2,1,0}-I_{3,1,0,1})\\
I_{Q,R,S,T}&=&\int_0^{\infty}\frac{\mathrm{d}s}{s}s^{-(b-a)}\frac{\left[1-\left(\frac{\eta}{s}\right)^2\right]^T\left[1-(\eta s)^2\right]^S}{\left[1+\left(\frac{\eta}{s}\right)^2\right]^{R+T}\left[1+(\eta s)^2\right]^{Q+R}}.
\eea
The repulsion force is effectively dominated by the behavior of the powers of $\eta$ appearing in the prefactors. As such, the repulsive force can be expected to decay exponentially in the distance $\delta$ between the two domain walls. This result is a direct consequence of the ultrathin film limit which allows one to write the magnetic stray field as an effective shape anisotropy and is not expected to qualitatively change for the case of 2-dimensional magnetic skyrmions considered in the main body of the paper.

%% file: Sections/appC-Table.tex
In this section we list all the material parameters explored throughout our micromagnetic simulations. In order to perform our analysis of the skyrmion-skyrmion repulsive forces in the main text, we required that the micromagnetically simulated magnetic textures relax to their stable values within short enough timescales such that all further dynamical behavior could be considered {\it physically relevant}. The seven sets of material values shown below are named {\it S1-S7} ({\it S2} corresponds to results in the main text).

\begin{table}[H]
\centering
\caption{Material Parameters}
\label{tab1}
\begin{tabular}{ccccccc}
\hline
\multicolumn{1}{|l|}{\cellcolor[HTML]{000000}} & \multicolumn{1}{c|}{$M_S$ ($M\mathrm{A}/\mathrm{m}$)} & \multicolumn{1}{c|}{$J_{ex}$ ($f\mathrm{J}/\mathrm{m}$)} & \multicolumn{1}{c|}{$D$ ($m\mathrm{J}/\mathrm{m}^2$)} & \multicolumn{1}{c|}{$K$ ($M\mathrm{J}/\mathrm{m}^3$)} & \multicolumn{1}{c|}{$B$ ($m\mathrm{T}$)} & \multicolumn{1}{c|}{$R_{\mathrm{skx}}$ ($n\mathrm{m}$)} \\ \hline
\multicolumn{1}{|c|}{S1}                       & \multicolumn{1}{c|}{1.4}       & \multicolumn{1}{c|}{27.5}       & \multicolumn{1}{c|}{2.05}      & \multicolumn{1}{c|}{1.45}      & \multicolumn{1}{c|}{0}      & \multicolumn{1}{c|}{23.2}      \\ \hline
\multicolumn{1}{|c|}{S2}                       & \multicolumn{1}{c|}{1.45}      & \multicolumn{1}{c|}{27.5}       & \multicolumn{1}{c|}{2.05}      & \multicolumn{1}{c|}{1.45}      & \multicolumn{1}{c|}{20}     & \multicolumn{1}{c|}{19.1}      \\ \hline
\multicolumn{1}{|c|}{S3}                       & \multicolumn{1}{c|}{1.4}       & \multicolumn{1}{c|}{27.5}       & \multicolumn{1}{c|}{2.6}       & \multicolumn{1}{c|}{1.45}      & \multicolumn{1}{c|}{20}     & \multicolumn{1}{c|}{30.3}      \\ \hline
\multicolumn{1}{|c|}{S4}                       & \multicolumn{1}{c|}{1.4}       & \multicolumn{1}{c|}{27.5}       & \multicolumn{1}{c|}{2.6}       & \multicolumn{1}{c|}{1.60}      & \multicolumn{1}{c|}{-15}    & \multicolumn{1}{c|}{14.9}      \\ \hline
\multicolumn{1}{|c|}{S5}                       & \multicolumn{1}{c|}{1.5}       & \multicolumn{1}{c|}{25.0}       & \multicolumn{1}{c|}{2.4}       & \multicolumn{1}{c|}{1.60}      & \multicolumn{1}{c|}{30}     & \multicolumn{1}{c|}{27.9}      \\ \hline
\multicolumn{1}{|c|}{S6}                       & \multicolumn{1}{c|}{1.0}       & \multicolumn{1}{c|}{5.0}        & \multicolumn{1}{c|}{2.05}      & \multicolumn{1}{c|}{0.9}       & \multicolumn{1}{c|}{150}    & \multicolumn{1}{c|}{11.9}      \\ \hline
\multicolumn{1}{|c|}{S7}                       & \multicolumn{1}{c|}{1.2}       & \multicolumn{1}{c|}{8.0}        & \multicolumn{1}{c|}{2.1}       & \multicolumn{1}{c|}{0.9}       & \multicolumn{1}{c|}{150}    & \multicolumn{1}{|c|}{14.9}   \\ \hline  
\end{tabular}
\end{table} 

\begin{table}[H]
\centering
\caption{Skx-Skx Repulsion}
\label{tab2}
\begin{tabular}{cccc}
\hline
\multicolumn{1}{|l|}{\cellcolor[HTML]{000000}} & \multicolumn{3}{|c|}{$|F|\propto \exp\left\{-\frac{a_2d^2+a_1d+a_0}{1+d}\right\}$}                                                                \\ \hline
\multicolumn{1}{|l|}{\cellcolor[HTML]{000000}} & \multicolumn{1}{c|}{$a_2$}     & \multicolumn{1}{c|}{$a_1$}     & \multicolumn{1}{c|}{$a_0$}      \\ \hline
\multicolumn{1}{|c|}{S1}                       & \multicolumn{1}{c|}{1.245}  & \multicolumn{1}{c|}{1.249}  & \multicolumn{1}{c|}{-3.449}  \\ \hline
\multicolumn{1}{|c|}{S2}                       & \multicolumn{1}{c|}{2.709}  & \multicolumn{1}{c|}{-5.643} & \multicolumn{1}{c|}{0.964}   \\ \hline
\multicolumn{1}{|c|}{S3}                       & \multicolumn{1}{c|}{2.488}  & \multicolumn{1}{c|}{-1.899} & \multicolumn{1}{c|}{-2.676}  \\ \hline
\multicolumn{1}{|c|}{S4}                       & \multicolumn{1}{c|}{1.13}   & \multicolumn{1}{c|}{2.171}  & \multicolumn{1}{c|}{-6.623}  \\ \hline
\multicolumn{1}{|c|}{S5}                       & \multicolumn{1}{c|}{2.219}  & \multicolumn{1}{c|}{-1.159} & \multicolumn{1}{c|}{-3.358}  \\ \hline
\multicolumn{1}{|c|}{S6}                       & \multicolumn{1}{c|}{0.059} & \multicolumn{1}{c|}{10.757} & \multicolumn{1}{c|}{-16.738} \\ \hline
\multicolumn{1}{|c|}{S7}                       & \multicolumn{1}{c|}{0.766}  & \multicolumn{1}{c|}{5.452}  & \multicolumn{1}{c|}{-11.894} \\ \hline
\end{tabular}
\end{table}

\begin{figure}[H]
	\centerline{\includegraphics[width=4in]{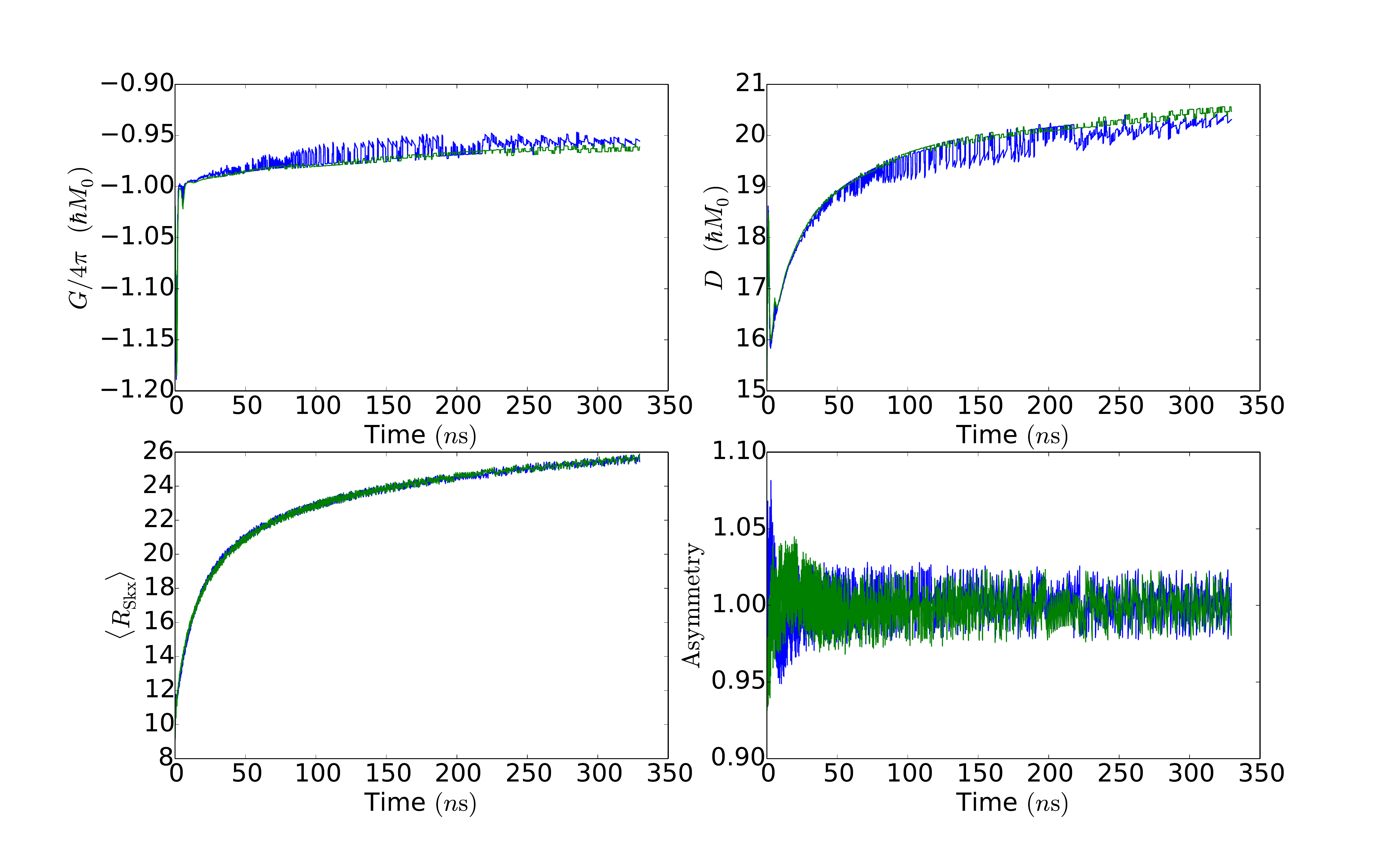}}
	\caption{{\footnotesize Topological charge, gyrodamping, average radius and asymmetry observed through micromagnetic simulations of two repelling skyrmions via material parameter set S1.}}
	\label{R1}
\end{figure}

\begin{figure}[H]
	\centerline{\includegraphics[width=4in]{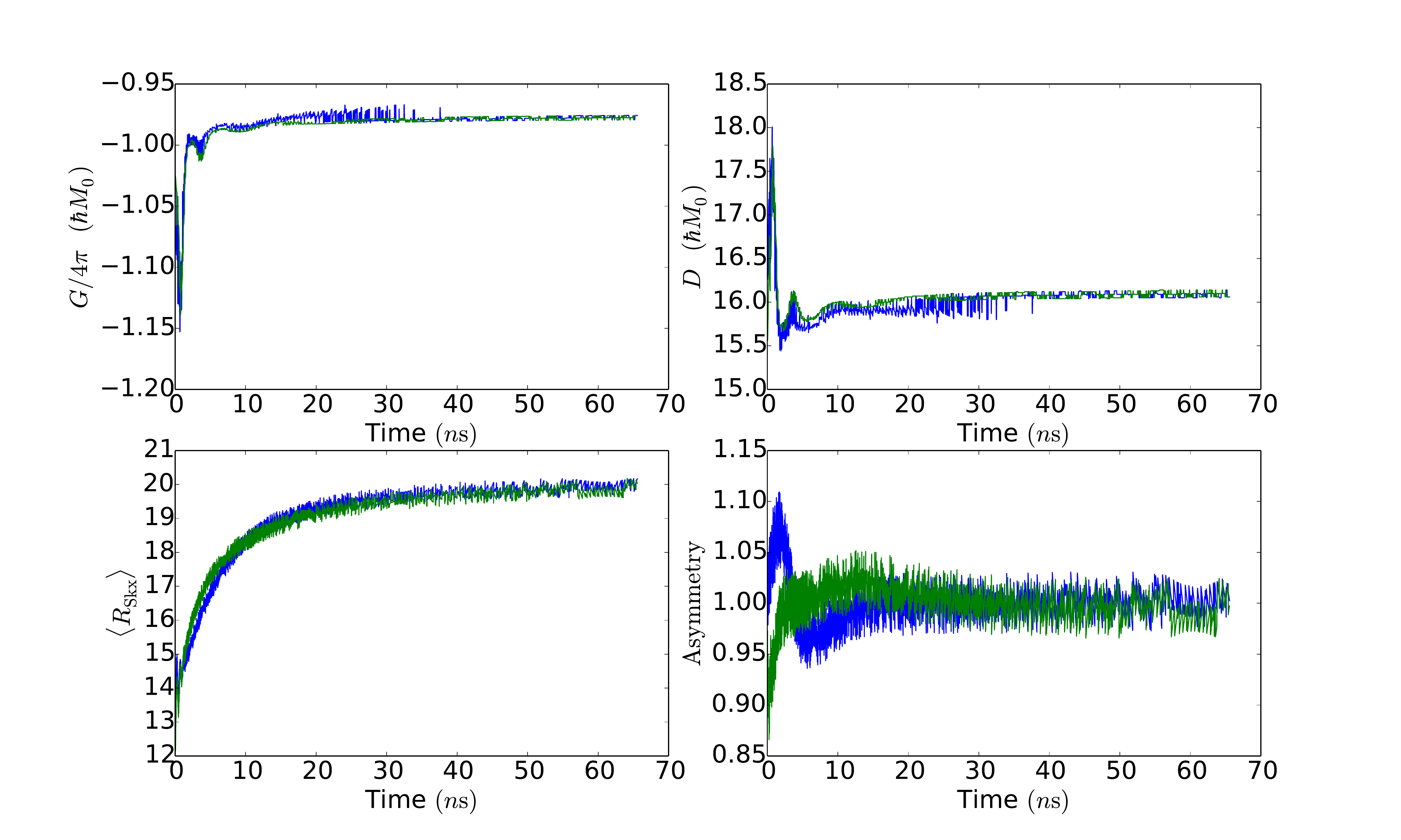}}
	\caption{{\footnotesize Topological charge, gyrodamping, average radius and asymmetry observed through micromagnetic simulations of two repelling skyrmions via material parameter set S2.}}
	\label{R2}
\end{figure}

\begin{figure}[H]
	\centerline{\includegraphics[width=4in]{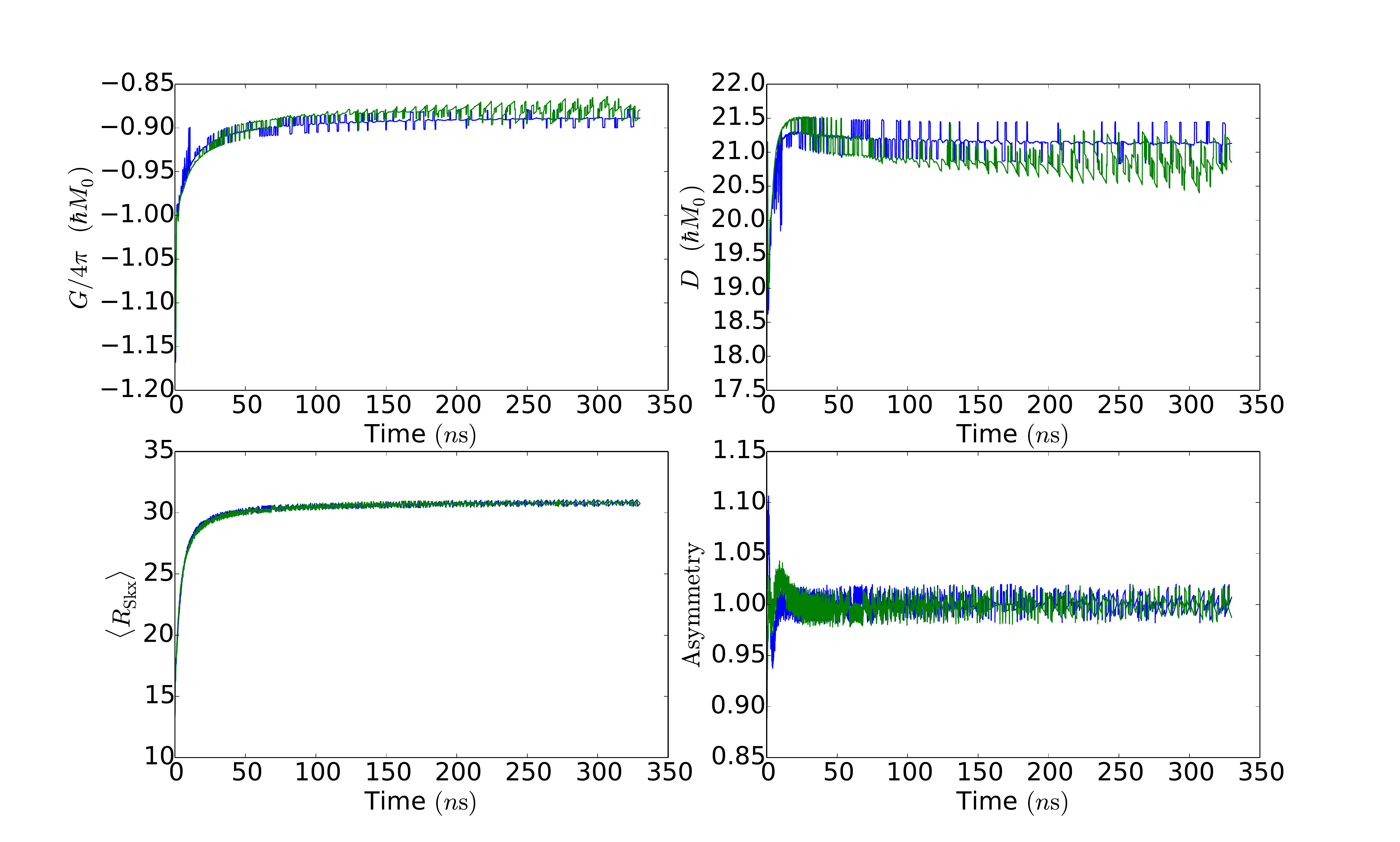}}
	\caption{{\footnotesize Topological charge, gyrodamping, average radius and asymmetry observed through micromagnetic simulations of two repelling skyrmions via material parameter set S3.}}
	\label{R3}
\end{figure}

\begin{figure}[H]
	\centerline{\includegraphics[width=4in]{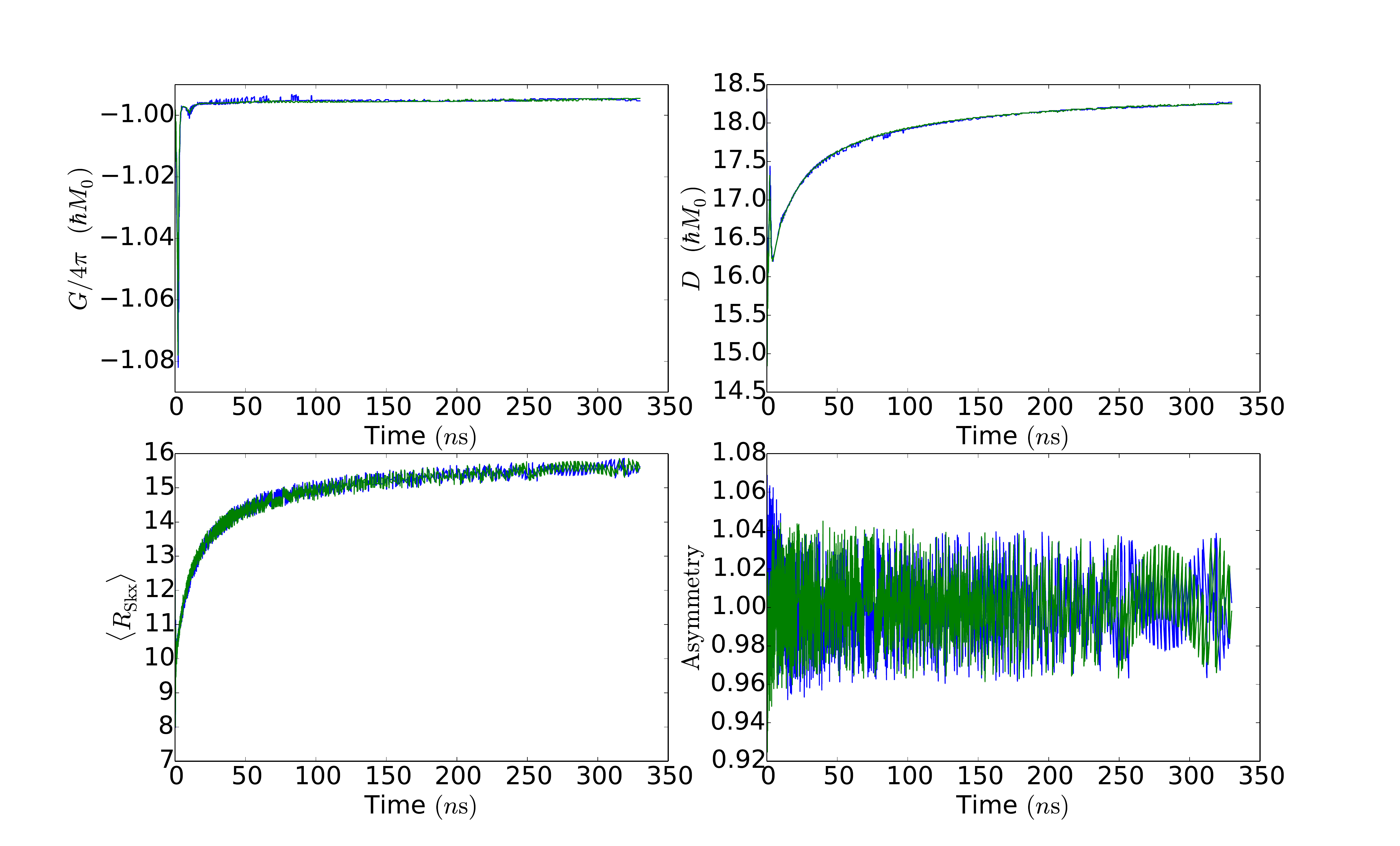}}
	\caption{{\footnotesize Topological charge, gyrodamping, average radius and asymmetry observed through micromagnetic simulations of two repelling skyrmions via material parameter set S4.}}
	\label{R4}
\end{figure}

\begin{figure}[H]
	\centerline{\includegraphics[width=4in]{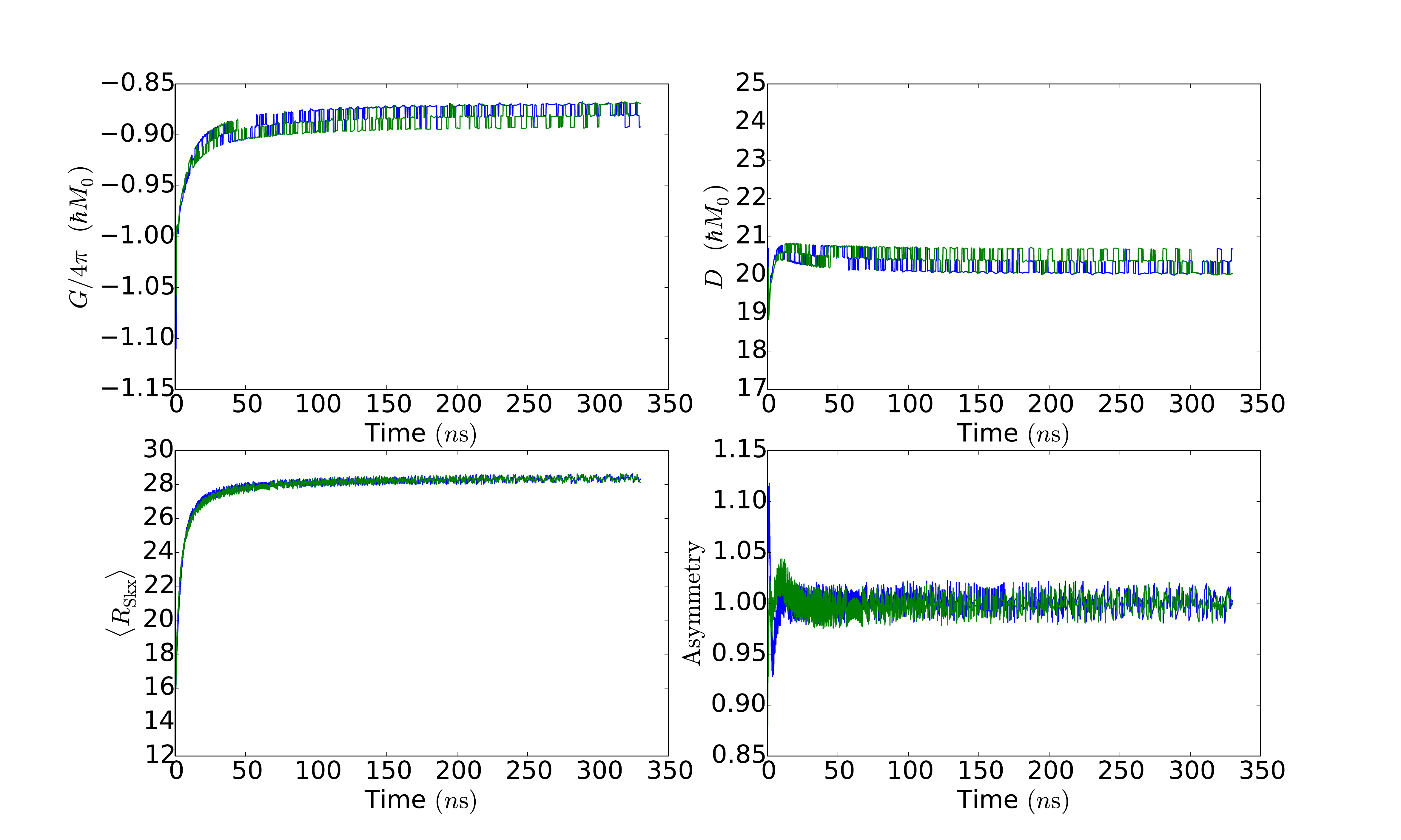}}
	\caption{{\footnotesize Topological charge, gyrodamping, average radius and asymmetry observed through micromagnetic simulations of two repelling skyrmions via material parameter set S5.}}
	\label{R5}
\end{figure}

\begin{figure}[H]
	\centerline{\includegraphics[width=4in]{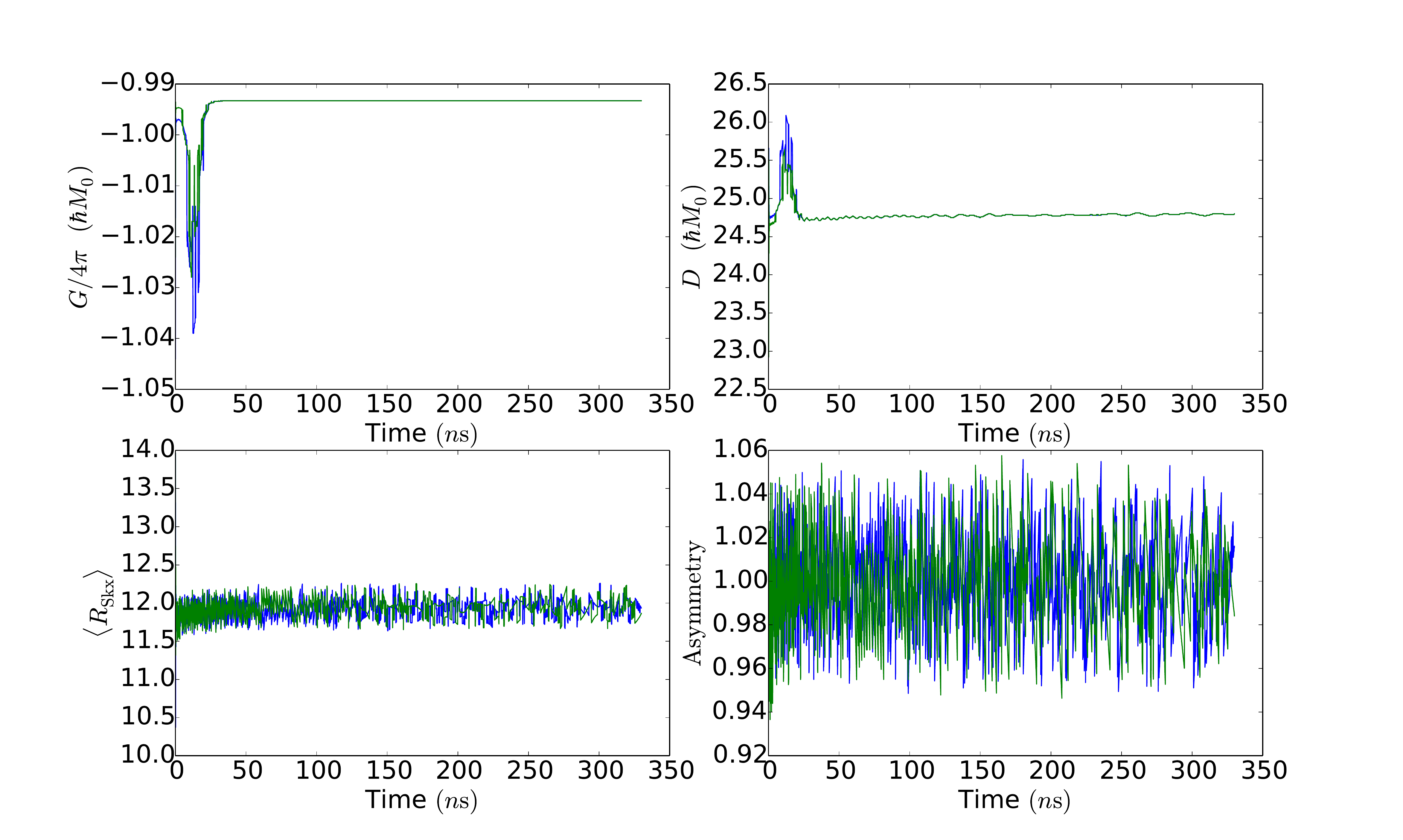}}
	\caption{{\footnotesize Topological charge, gyrodamping, average radius and asymmetry observed through micromagnetic simulations of two repelling skyrmions via material parameter set S6.}}
	\label{R6}
\end{figure}

\begin{figure}[H]
	\centerline{\includegraphics[width=4in]{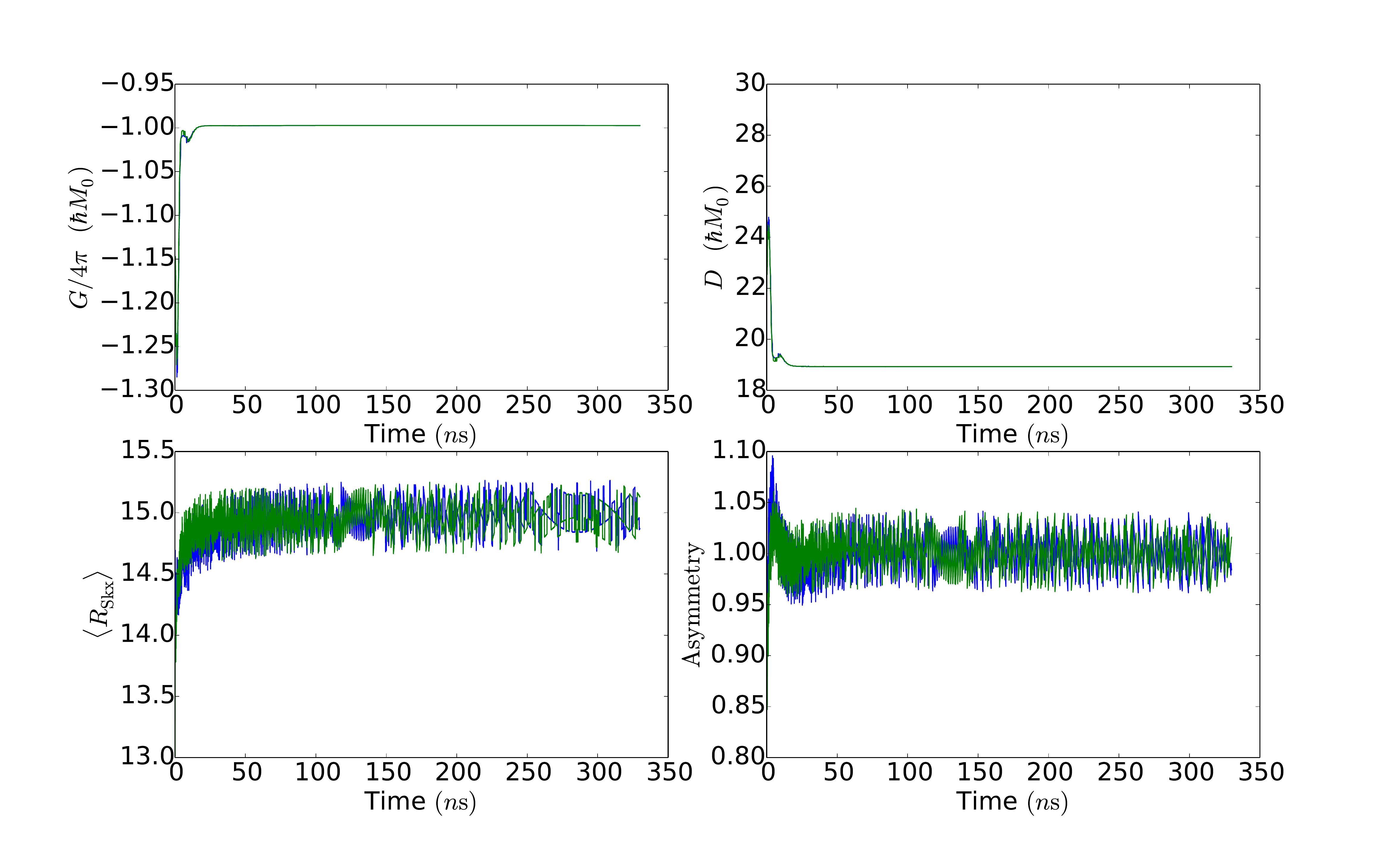}}
	\caption{{\footnotesize Topological charge, gyrodamping, average radius and asymmetry observed through micromagnetic simulations of two repelling skyrmions via material parameter set S7.}}
	\label{R7}
\end{figure}